\documentclass[aps,pre,preprint,nofootinbib,superscriptaddress,showpacs]{revtex4-1}
\pdfoutput=1
\usepackage{subfigure}
\usepackage{amsmath,graphicx,amssymb}
\usepackage{multirow,color}
\usepackage{bm}
\usepackage{verbatim,url}

\renewcommand{\vec}[1]{\bm{#1}}
\newcommand{\etal}{\textit{et al.}}
\newcommand{\Ceps}{C_\varepsilon}

\newcommand{\vep}{\varepsilon}

\newcommand{\Rl}{R_{\lambda}}

\newcommand{\beq}{\begin{equation}}
\newcommand{\eeq}{\end{equation}}
\begin{document}

\title{Spectral analysis of structure functions and their scaling exponents in forced isotropic 
turbulence}

\author{W.~D. McComb}
\email{wdm@ph.ed.ac.uk}
 \affiliation{
 SUPA, School of Physics and Astronomy,
 University of Edinburgh, James Clerk Maxwell Building, The King's Buildings, Edinburgh EH9 3JZ, UK}
\author{S.~R. Yoffe}
 \affiliation{
 SUPA, Department of Physics,
 University of Strathclyde, John Anderson Building, 107 Rottenrow East, Glasgow G4 0NG, UK}
 \author{M.~F. Linkmann}
 \affiliation{
 SUPA, School of Physics and Astronomy,
 University of Edinburgh, James Clerk Maxwell Building, The King's Buildings, Edinburgh EH9 3JZ, UK}
 \author{A. Berera}
 \affiliation{
 SUPA, School of Physics and Astronomy,
 University of Edinburgh, James Clerk Maxwell Building, The King's Buildings, Edinburgh EH9 3JZ, UK}


\begin{abstract}
The pseudospectral method, in conjunction with a 
new technique for obtaining scaling exponents $\zeta_n$ from the
structure functions $S_n(r)$, is presented as an alternative to 
the extended self-similarity (ESS) method and the use of generalized 
structure functions. We propose plotting the ratio 
$|S_n(r)/S_3(r)|$ against the separation $r$ in accordance with a 
standard technique for analyzing experimental data.
This method differs from the ESS technique,
which plots $S_n(r)$ against $S_3(r)$, with the assumption $S_3(r) \sim r$. Using our
method for the particular case of $S_2(r)$ we obtain the new result
that the exponent $\zeta_2$ decreases as the Taylor-Reynolds number
increases, with  $\zeta_2 \to 0.679 \pm 0.013$ as $R_\lambda \to \infty$. This
supports the idea of finite-viscosity corrections to the K41 prediction
for $S_2$, and is the opposite of the result obtained by ESS. 
The pseudospectral method also permits the forcing 
to be taken into account exactly through the calculation of the 
energy input in real space from the work spectrum of the stirring
forces.  
\end{abstract}


\maketitle

\newpage

\section{Introduction}
In this paper we revisit an old, but unresolved, issue in turbulence:
the controversy that continues to surround the Kolmogorov theory (or
K41) \cite{Kolmogorov41a,Kolmogorov41b}. This controversy began  with
the publication in 1962 of Kolmogorov's `refinement of previous
hypotheses' (K62), which gave a role to the intermittency of the dissipation
rate \cite{Kolmogorov62}. From this beginning, the search for
`intermittency corrections' has grown into a veritable industry over the
years: for a general discussion, see the book by Frisch \cite{Frisch95} and the
review by Boffetta, Mazzino and Vulpiani \cite{Boffetta08}. The term 
`intermittency corrections' is rather
tendentious, as no relationship has ever been demonstrated between
intermittency, which is a  property of a single realization, and the
ensemble-averaged energy fluxes which underlie K41, and it is now
increasingly replaced by `anomalous exponents'. It has also been
observed by Kraichnan \cite{Kraichnan74}, Saffman \cite{Saffman77}, Sreenivasan 
\cite{Sreenivasan99} and Qian \cite{Qian00}
 that the title of K62 is misleading. It in fact represents a profoundly
different view of the underlying physics of turbulence, as compared to
K41. For this reason alone it is important to resolve this controversy.

While this search has been a dominant theme in turbulence for many
decades, at the same time there has been a small but significant number
of theoretical papers exploring the effect of finite Reynolds numbers on
the Kolmogorov exponents, such as the work by Effinger and Grossmann
\cite{Effinger87}, Barenblatt and Chorin \cite{Barenblatt98a}, 
Qian \cite{Qian00}, Gamard and George \cite{Gamard00} and Lundgren \cite{Lundgren02}.
 All of these papers have something to say; but the last one is perhaps
the most compelling, as it appears to offer a rigorous proof of the
validity of K41 in the limit of infinite Reynolds
number. This is reinforced by the author's comparison with the experimental
results of Mydlarski and Warhaft \cite{Mydlarski96}.

The controversy surrounding K41 basically amounts to: `intermittency
corrections' \emph{versus} `finite Reynolds number effects'. The former
are expected to increase with increasing Reynolds number, the latter to
decrease. In time, direct numerical simulation (DNS)
should establish the nature of high-Reynolds-number asymptotics, and so
decide between the two. In the meantime, one would like to find some way
of extracting the `signature' of this information from current simulations.

As is well known, one way of doing this is by Extended Self Similarity (ESS). 
The study of turbulence structure functions (\emph{e.g.}~see van Atta and Chen
\cite{vanatta70} and Anselmet, Gagne, Hopfinger and Antonia \cite{Anselmet84}) 
was transformed in the mid-1990s by the
introduction of ESS by Benzi and co-workers \cite{Benzi93,Benzi95}. Their
method of plotting results for $S_n(r)$ against $S_3(r)$, rather than
against the separation $r$, showed extended regions of apparent scaling
behavior even at low-to-moderate values of the Reynolds number, and was
widely taken up by others \emph{e.g.}~Fukayama, Oyamada, Nakano, Gotoh and Yamamoto
\cite{Fukayama00}, Stolovitzky, Sreenivasan and Juneja \cite{Stolovitzky93}, 
Meneveau \cite{Meneveau96}, Grossmann, Lohse and Rech \cite{Grossmann97} and 
Sain, Manu and Pandit \cite{Sain98}.
A key feature of this work was the implication that corrections to the
exponents of structure functions increase with increasing Reynolds number,
which suggests that intermittency is the dominant effect. In the next
section, we will explain ESS in more detail, in terms of how it relates
to other ways of obtaining exponents.

\section{Structure functions, exponents and ESS}
In this section we discuss the various ways in which exponents are
defined and measured.
The longitudinal structure functions are defined as
\begin{equation}
 \label{eq:SF}
 S_n(r) = \left\langle \delta u^n_L(r) \right\rangle \ ,
\end{equation}
where the (longitudinal) velocity increment is given by
\begin{equation}
 \delta u_L(r) = \Big[ \vec{u}(\vec{x}+\vec{r},t) - \vec{u}(\vec{x},t) \Big]
 \cdot \vec{\hat{r}} \ .
\end{equation}
Integration of the K\'arm\'an-Howarth equation (KHE)
leads, \emph{in the limit of infinite Reynolds number}, to the
Kolmogorov  `4/5' law, $S_3(r) = -(4/5)\varepsilon r$. If the
$S_n$, for $n \geq 4$, exhibit a range of power-law behavior, then; in general, and
solely  on dimensional grounds, the structure functions of order $n$ are
expected to  take the form
\begin{equation}
 S_n(r) = C_n (\varepsilon r)^{n/3} \ .
\end{equation}

Measurement of the structure functions has repeatedly found a deviation
from the above dimensional prediction for the exponents. If the structure
functions are taken to scale with exponents $\zeta_n$, thus:
\begin{equation}
 S_n(r) \sim r^{\zeta_n} \ ,
\end{equation}
then it has been found \cite{Anselmet84,Benzi95} that the difference
$\Delta_n = \lvert n/3 - \zeta_n\rvert$ is non-zero and increases with order
$n$.  Exponents $\zeta_n$ which differ from $n/3$ are often referred to as
\emph{anomalous exponents} \cite{Benzi95}.

\begin{figure}[tbp]
 \begin{center}
  \includegraphics[width=0.8\textwidth]{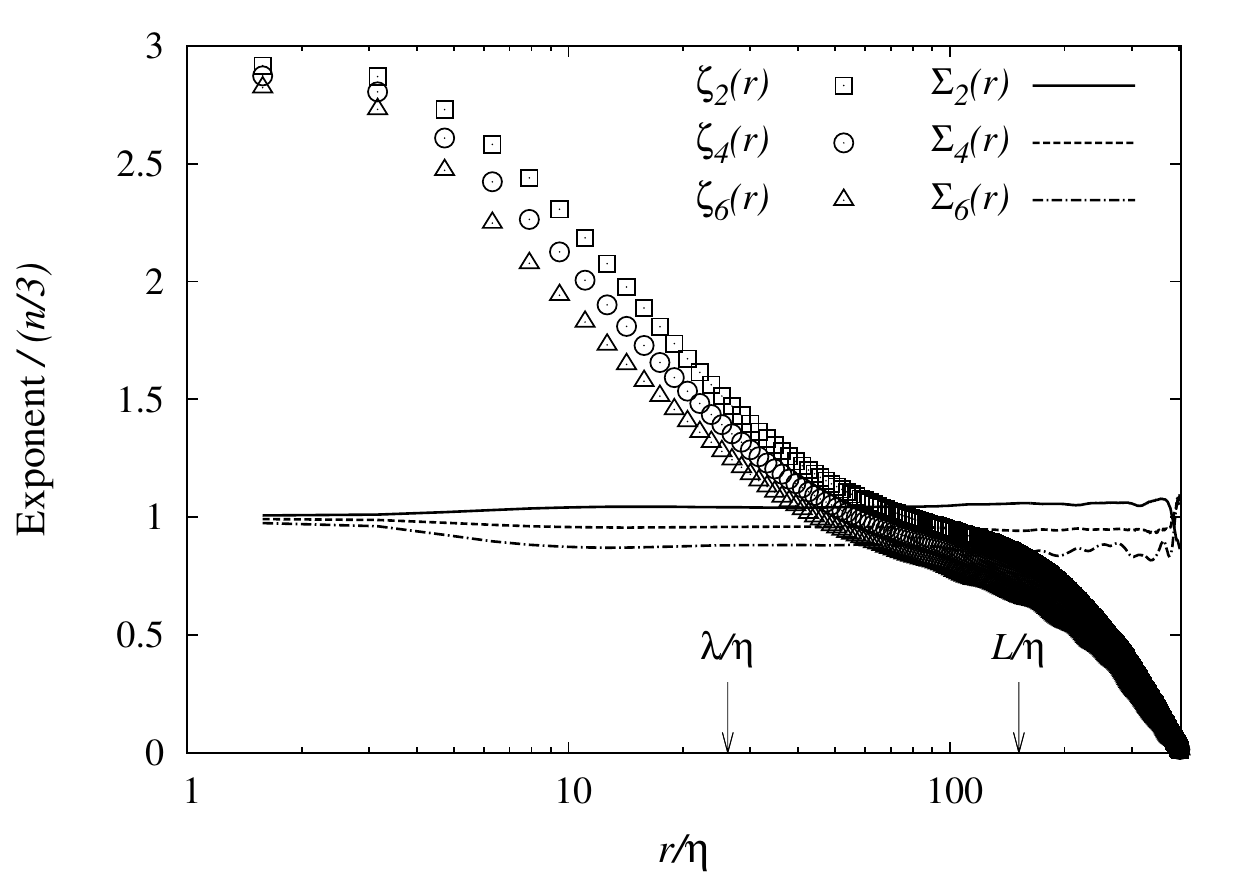}
 \end{center}
 \caption{Comparison of our values of local-slope exponents $\zeta_n(r)$ 
(symbols) given in \eqref{eq:local_zeta} with our values of the ESS exponents 
$\Sigma_n(r)$ (lines) as calculated from equation \eqref{eq:sigma}, divided by $n/3$ to show their 
relationship to K41 exponents. Results are given for $n=2, 4, 6$. 
Both sets of exponents were calculated from the real-space velocity field and are
 presented here for $R_\lambda = 177$.  The separation, $r$, has been scaled
 on the dissipation scale, $\eta = (\nu_0^3/\varepsilon)^{1/4}$.}
 \label{fig:local_grads}
\end{figure}

In order to study the behavior of the exponents $\zeta_n$, it is usual
to make a log-log plot of $S_n$ against $r$, and measure the local slope:
\begin{equation}
 \label{eq:local_zeta}
 \zeta_n(r) = \frac{d\log{S_n(r)}}{d\log{r}} \ .
\end{equation}
Following Fukayama \etal\ \cite{Fukayama00}, the presence of a
plateau when any $\zeta_n(r)$ is plotted against $r$ indicates a constant
exponent, and hence a scaling region.
Yet, it  is not until comparatively high Reynolds numbers are
attained that such a  plateau is found. Instead, as seen in Fig.
\ref{fig:local_grads} (symbols),  even for the relatively large value of
Reynolds number, $R_\lambda = 177$, a scaling region cannot be
identified. (We note that Grossmann \etal\ \cite{Grossmann97} have
argued that a
\emph{minimum} value of $R_\lambda \sim 500$ is needed for satisfactory
direct
measurement of local scaling exponents.)

The introduction of ESS relied on the fact that $S_3$ scales with
$\zeta_3 =  1$ in the inertial range. Benzi \etal\ \cite{Benzi93}
argued that if 
\begin{equation}
 S_n(r) \sim [S_3(r)]^{\zeta^*_n}  \ , \quad\textrm{with}\quad \zeta^*_n = 
 \zeta_n/\zeta_3 \ .
\end{equation}
$\zeta^*_n$ should then be equivalent to $\zeta_n$ in the scaling region.

A practical difficulty led to a further step.  The statistical
convergence of odd-order structure functions is  significantly slower
than that for even-orders, due to the delicate balance  of positive and
negative values involved in the former \cite{Fukayama00}. To  overcome
this, \emph{generalized} structure functions, where the velocity
difference is replaced by its modulus,
have been introduced
\cite{Benzi93} (see also \cite{Stolovitzky93,Fukayama00}),
\begin{equation}
G_n(r) = \left\langle \lvert \delta u_L(r) \rvert^n \right\rangle \sim
r^{\zeta'_n},
\end{equation}
with scaling exponents $\zeta'_n$.  The fact that $S_3 \sim r$ in the
inertial range does not rigorously imply that $G_3 \sim r$ in the same
range. But, by plotting $G_3(r)$ against $\lvert S_3(r)\rvert$, Benzi \etal\
\cite{Benzi95} showed that, for $R_\lambda = 225$ -- 800, the third-order
exponents satisfied $\zeta'_3 \simeq 1.006\zeta_3$.  Hence, it is now
generally assumed that $\zeta'_n$ and $\zeta_n$ are equal
(although, Fig.~2 in Belin, Tabeling and Willaime \cite{Belin96} implies
some discrepancy at the largest length scales, and the authors note that the
exponents $\zeta'_n$ and $\zeta_n$ need not be the same).
Thus, by extension, $G_3$ with $\zeta'_3 = 1$, leads to
\begin{equation}
\label{eq:sigma}
G_n(r) \sim [G_3(r)]^{\Sigma_n}, \quad\textrm{with}\quad \Sigma_n = 
\zeta'_n/\zeta'_3 \ .
\end{equation}
Benzi \etal\ \cite{Benzi93} found
that plotting their results on  this basis gave a larger scaling region.
This extended well into the dissipative lengthscales and allowed
exponents to be more easily extracted from the data. Also, Grossmann
\etal\ \cite{Grossmann97} state that the use of generalized structure
functions is
essential to take full advantage of ESS.


There is, however, an alternative to the use of generalized structure
functions. This is the \emph{pseudospectral method}.
In using this for some of the present work, we followed the example of
Qian \cite{Qian97,Qian99}, Tchoufag, Sagaut and Cambon
\cite{Tchoufag12} and Bos, Chevillard, Scott and Rubinstein \cite{Bos12}, 
who obtained $S_2$ and $S_3$ from the energy and
energy transfer spectra, respectively, by means of exact quadratures.

The organization of our own work in this paper is now as follows. 
We begin with a description of our DNS before illustrating ESS, 
using results from our own simulations, in section \ref{sec:ESS}, 
where we show that our results for ESS agree closely with those 
of other investigations \cite{Benzi95,Fukayama00}. These
particular results were obtained in the usual way by direct 
convolution sums, using a statistical ensemble, and the generalized 
structure functions. In section \ref{sec:spectral} we describe 
the theoretical basis for using the pseudospectral method 
\cite{Qian97,Qian99,Tchoufag12,Bos12} which includes a rigorous 
derivation of the forcing in the real-space energy balance equation. 
This is followed by the introduction of a new scaling exponent in section 
\ref{sec:new_exponent} and a presentation of our numerical results
for seven Taylor-scale Reynolds numbers spanning the range 
$101.3 \leq R_\lambda \leq 435.2$. 

\section{Numerical method}

We used a pseudospectral DNS, with full dealiasing implemented by
truncation of the velocity field according to the two-thirds rule \cite{Orszag71}. Time
advancement for the viscous term was performed exactly using an integrating
factor, while the non-linear term was stepped forward in time using Heun's method
\cite{Heun00},
which is a second-order predictor-corrector routine. Each
simulation was started from a Gaussian-distributed random field with a
specified energy spectrum, which followed $k^4$ for the low-$k$
modes. Measurements were taken after the simulations had reached a
stationary state.
The system was forced by negative damping, with the Fourier transform
of the force $\vec{f}$ given by
\begin{align}
 \vec{f}(\vec{k},t) &=
      (\varepsilon_W/2 E_f) \vec{u}(\vec{k},t) \quad
\text{for} \quad  0 < \lvert\vec{k}\rvert < k_f ; \nonumber \\
  &= 0   \quad \textrm{otherwise},
\label{forcing}
\end{align}
where $\vec{u}(\vec{k},t)$ is the instantaneous velocity field (in
wavenumber space). The highest forced wavenumber, $k_f$, was chosen to
be $k_f = 2.5k_{min}$, where $k_{min}=2\pi/L_{box}=1$ is the lowest 
resolved wavenumber. As $E_f$ was the total energy contained in the forcing
band, this ensured that the energy injection rate was $\varepsilon_W =
\textrm{constant}$. It is worth noting that any method of energy
injection  employed in the numerical simulation of isotropic turbulence
is not experimentally realizable. The present method  of negative
damping has also been used in other investigations
\cite{Jimenez93,Yamazaki02,Kaneda03,Kaneda06}, albeit not necessarily
such that $\vep_W$ is maintained constant (although note the
theoretical analysis of this type of forcing by Doering and Petrov
\cite{Doering05}). Also, note that the correlation between the force and the
velocity is restricted to the very lowest wavenumbers.

For each Reynolds number studied, we used the same initial spectrum and
input rate $\varepsilon_W$. The only initial condition changed was the
value assigned to the (kinematic) viscosity. Once the initial transient
had passed, the velocity field was sampled every half a large-eddy
turnover time, $\tau = L/U$, where $L$ denotes the average integral scale and
$U$ the rms velocity.
The ensemble populated with these sampled
realizations was used, in conjunction with the usual shell averaging, to
calculate statistics. Simulations were run using lattices of size
$128^3,\ 256^3,\ 512^3,\ 1024^3$ and $2048^3$, with corresponding Reynolds
numbers ranging from $R_\lambda = 41.8$ up to $435.2$. The smallest
wavenumber was $k_\text{min} = 2\pi/L_\text{box} = 1$ in all
simulations, while the maximum wavenumber satisfied $k_\text{max}\eta
 \geqslant 1.30$ for all runs except one which satisfied
$k_\text{max}\eta \geqslant 1.01$, where $\eta$ is the Kolmogorov dissipation
lengthscale. The integral scale, $L$, was found to lie between
$0.23 L_{\text{box}}$ and $0.17 L_{\text{box}}$.
It can be seen in Figure 2 of McComb, Hunter and Johnston \cite{McComb01a} that a small-scale
resolution of $k_{max}\eta > 1.6$ is desirable in order to capture the relevant
dissipative physics. Evidently, this would restrict the attainable Reynolds
number of the simulated flow, and the reference suggests that
$k_{max}\eta \geqslant 1.3$ would still be acceptable (containing $\sim 99.5\%$
of dissipative dynamics \cite{Yoffe12}).
In contrast, at $k_{max}\eta \simeq 1$ a non-negligible part of dissipation
is not taken into account. Most high resolution DNSs of isotropic turbulence
 try to attain Reynolds numbers as high possible and thus opt for minimal
resolution requirements.
In this paper the simulations have been
conducted following a more conservative approach, where the emphasis has
been put on higher resolution, thus necessarily compromising to some extent
on Reynolds number.
Large-scale resolution has only relatively recently received attention in the
literature. As mentioned above, the largest scales of the flow are smaller
than a quarter of the simulation box size.
Details of the individual runs are summarized in Table \ref{tbl:simulations}.

Our simulations have been well validated by means of extensive and detailed
comparison with the results of other investigations. Further details of the 
performance of our code including verification of isotropy may be found in 
the thesis by Yoffe \cite{Yoffe12}, along with values for the 
Kolmogorov constant and velocity-derivative skewness; and a 
direct comparison with the freely-available pseudospectral
code {\tt{hit3d}} \cite{Schumakov07,Schumakov08}. 
Furthermore our data
reproduces the characteristic behavior for the plot of the dimensionless 
dissipation rate $\Ceps$ against $\Rl$
\cite{McComb14a}, and agree closely with other representative results in 
the literature, such as the work by Wang, Chen, Brasseur and Wyngaard 
\cite{Wang96}, Cao, Chen and Doolen \cite{Cao99}, Gotoh, Fukayama and Nakano 
\cite{Gotoh02}, Kaneda, Ishihara, Yokokawa, Itakura and Uno 
\cite{Kaneda03}, Donzis, Sreenivasan and Yeung \cite{Donzis05} and Yeung, 
Donzis and Sreenivasan \cite{Yeung12}, 
although there are some differences in forcing methods.

\begin{table}[tb!]
 \begin{center}
  \begin{ruledtabular}
  \begin{tabular}{cccccccl}
  $R_\lambda$ & $\nu_0$ & $N$ & $\varepsilon$ & $U$ & $L/L_\text{box}$ &
  $k_\text{max}\eta$ & $M$ \\
  \hline
  42.5  & 0.01    & 128  & 0.094 & 0.581 & 0.23 & 2.34 & 101 \\
  64.2  & 0.005   & 128  & 0.099 & 0.607 & 0.21 & 1.37 & 101 \\
  101.3 & 0.002   & 256  & 0.099 & 0.607 & 0.19 & 1.41 & 101 \\
  113.3 & 0.0018  & 256  & 0.100 & 0.626 & 0.20 & 1.31 \\
  176.9 & 0.00072 & 512  & 0.102 & 0.626 & 0.19 & 1.31 & 15  \\
  203.7 & 0.0005  & 512  & 0.099 & 0.608 & 0.18 & 1.01 \\
  217.0 & 0.0005  & 1024 & 0.100 & 0.630 & 0.19 & 2.02  \\
  276.2 & 0.0003  & 1024 & 0.100 & 0.626 & 0.18 & 1.38 \\
  335.2 & 0.0002  & 1024 & 0.102 & 0.626 & 0.18 & 1.01 \\
  435.2 & 0.00011 & 2048 & 0.102 & 0.614 & 0.17 & 1.30 
  \end{tabular}
  \end{ruledtabular}
 \end{center}
 \caption{A summary of the main parameters of our numerical simulations. The 
 values quoted for the dissipation rate $\vep$, the rms velocity $U$ and the
 integral scale $L$ are ensemble- and shell-averaged mean values, where the 
ensembles  have been populated with snapshots of the steady-state velocity 
field taken every half large eddy turnover time. $M$ counts individual 
realizations used to calulate ESS exponents for those runs for which the 
ESS method has been performed, not the size of the ensemble used for 
calculating statistics.}
 \label{tbl:simulations}
\end{table}

\section{Real-space calculation of the structure functions and ESS}
\label{sec:ESS}
In order to calculate the structure functions in real space, we calculate the 
longitudinal correlation of one lattice site with all other sites
   \begin{equation}
    S_n(r) = \frac{1}{3N^3} \sum_{\vec{x}} \bigg[ \Big( 
    u_x(\vec{x}+r\vec{e}_x) - u_x(\vec{x}) \Big)^n 
    + \Big( u_y(\vec{x}+r\vec{e}_y) - u_y(\vec{x}) \Big)^n
    + \Big( u_z(\vec{x}+r\vec{e}_z) - u_z(\vec{x}) \Big)^n \bigg] \ ,
   \end{equation}
for each realization. The results are subsequently ensemble-averaged over 
many realizations.

\begin{figure}
 \begin{center}
  \includegraphics[width=0.45\textwidth]{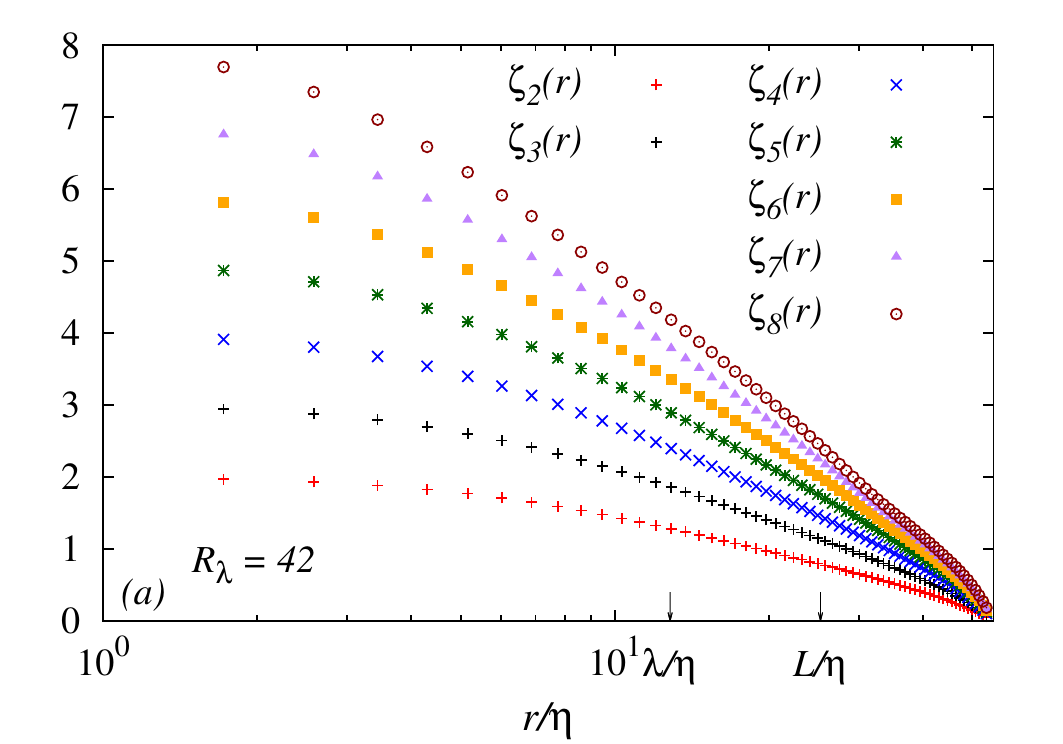}
  \includegraphics[width=0.45\textwidth]{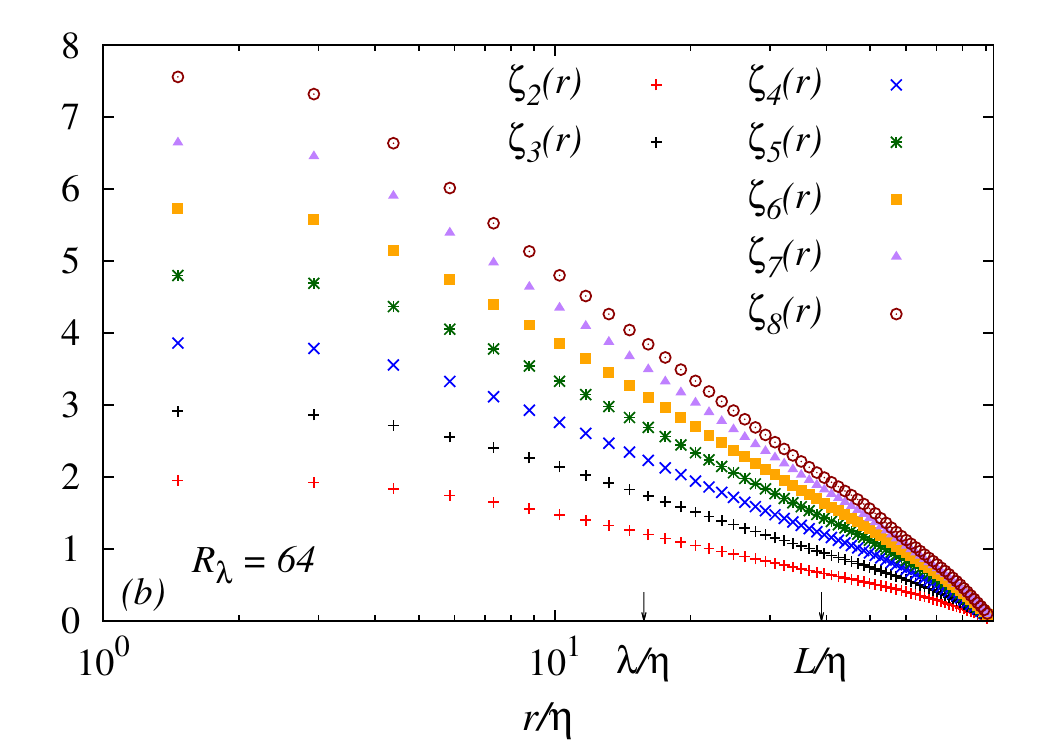}
  \includegraphics[width=0.45\textwidth]{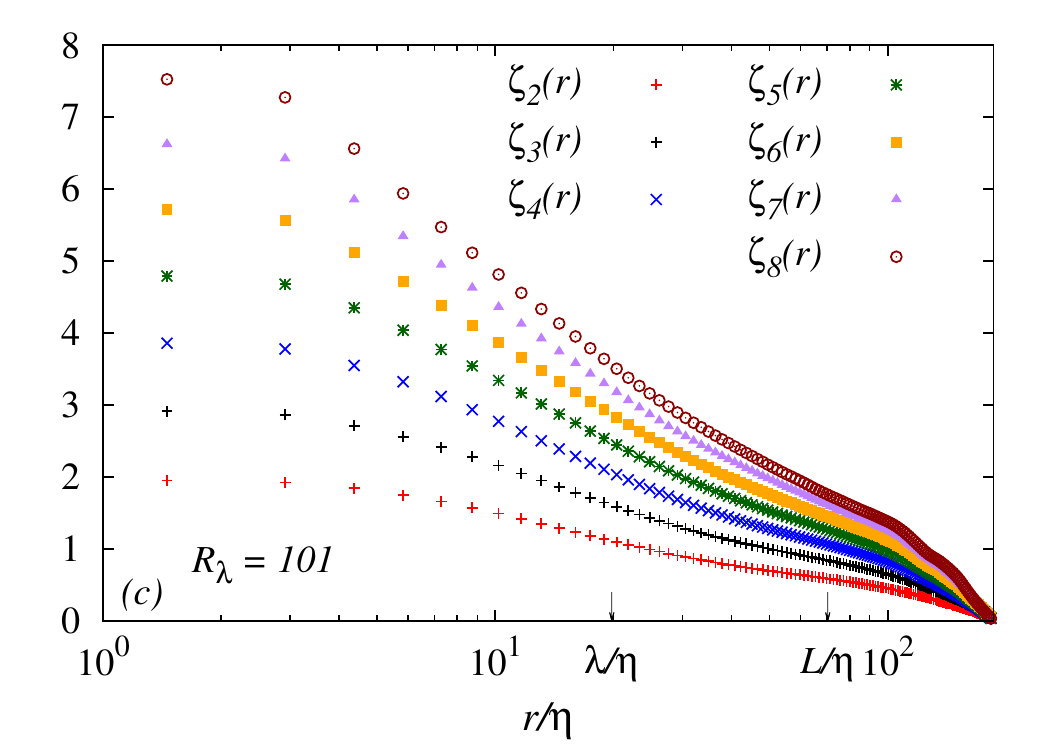}
  \includegraphics[width=0.45\textwidth]{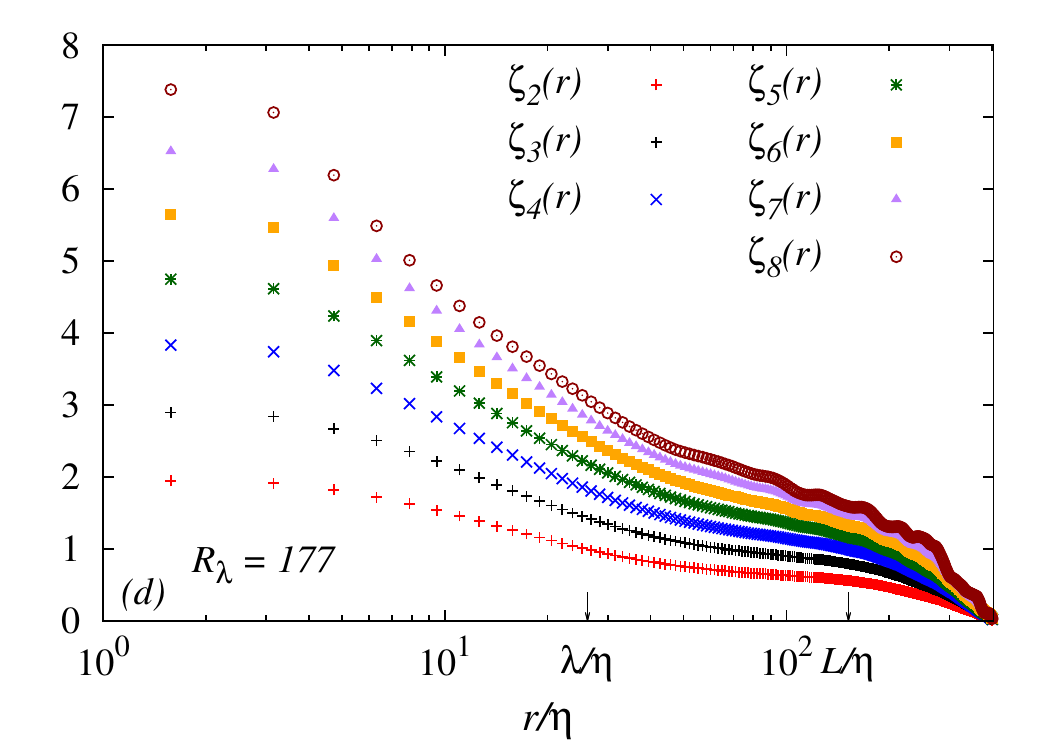}
 \end{center}
 \caption{(Color online) The local slopes $\zeta_n(r)$ of the structure functions $S_n(r)$ 
 for $n = 2,\cdots, 8$ from our DNS. The slopes are calculated using the structure functions 
 evaluated in real space for Reynolds numbers $\Rl=42$, 64, 101 and 177. Note the absence of 
 a scaling region in all cases.}
 \label{fig:local_slopes}
\end{figure}

\begin{figure}
 \begin{center}
  \includegraphics[width=0.45\textwidth]{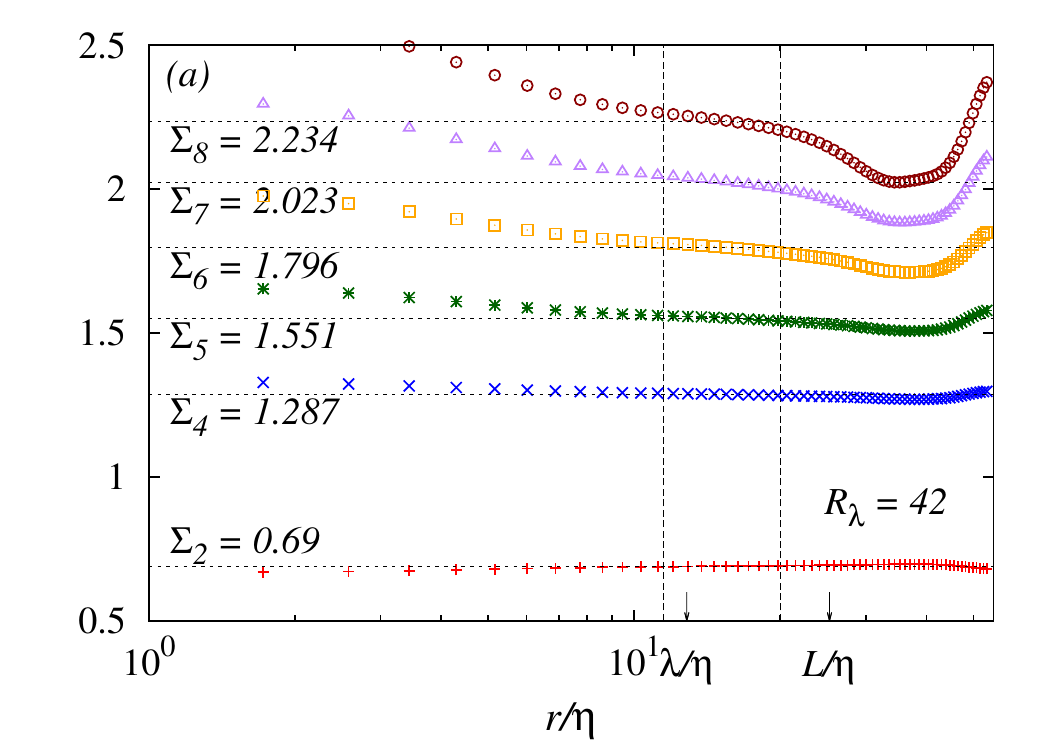}
  \includegraphics[width=0.45\textwidth]{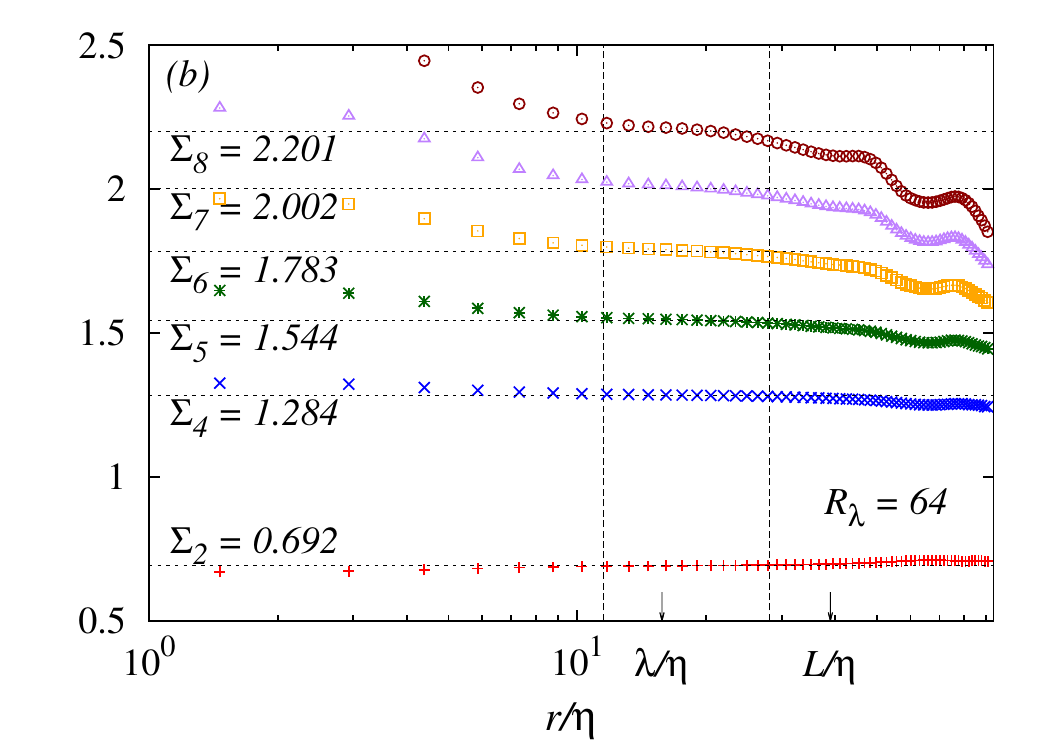}
  \includegraphics[width=0.45\textwidth]{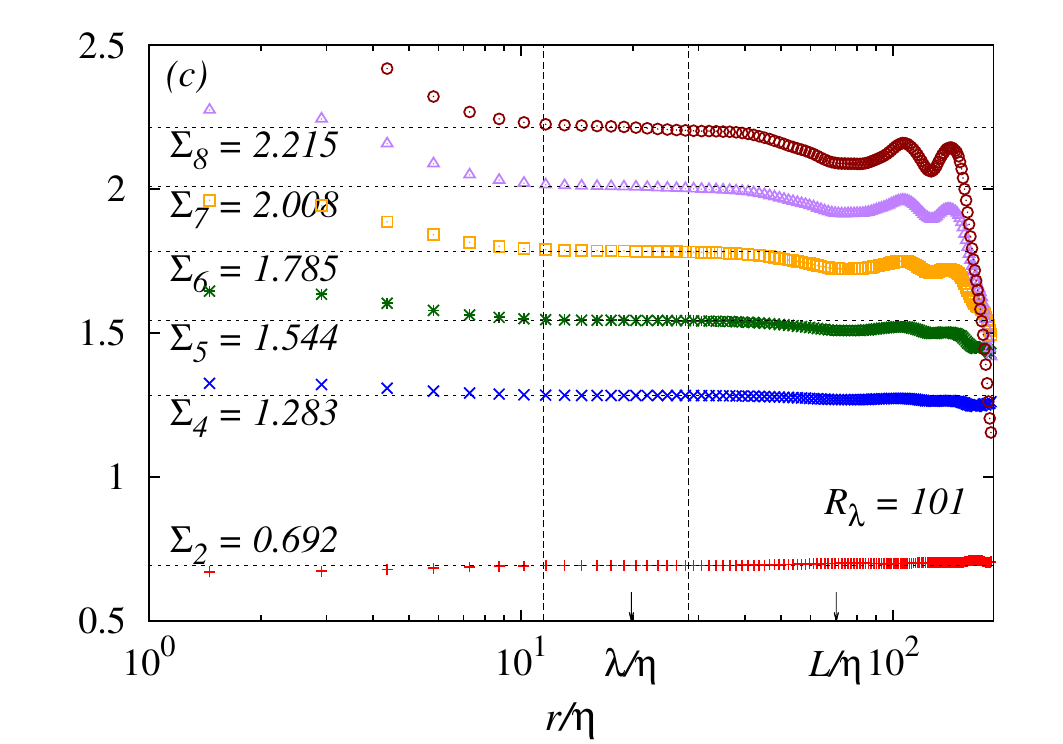}
  \includegraphics[width=0.45\textwidth]{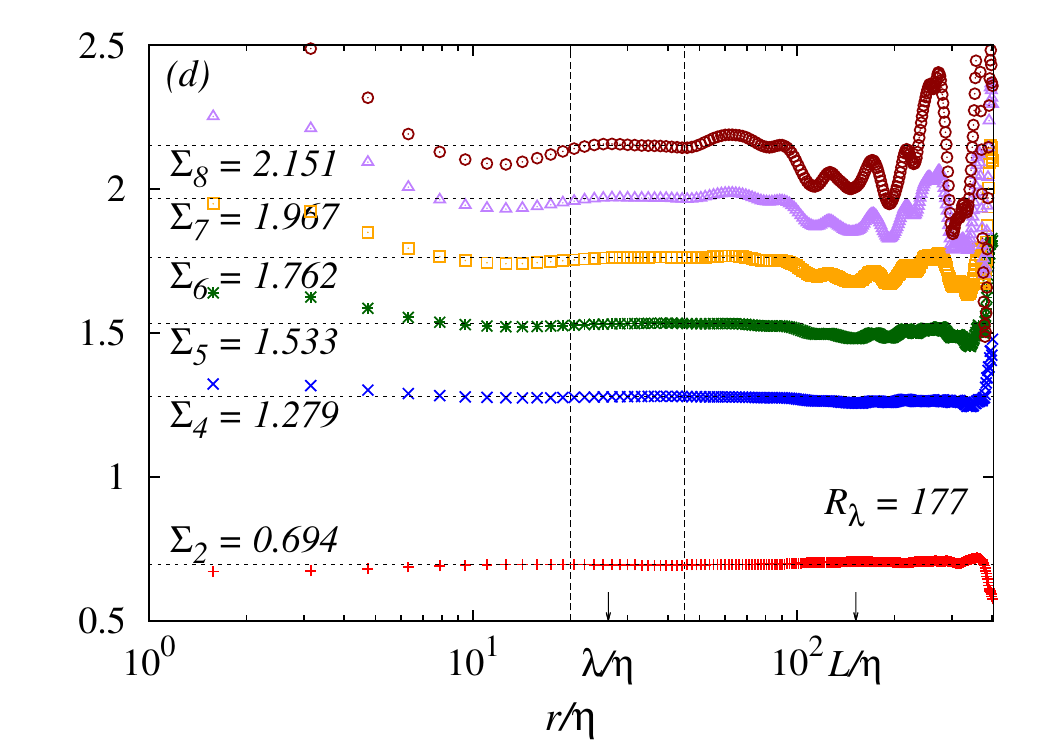}
 \end{center}
 \caption{(Color online) The same results as in Fig.~\ref{fig:local_slopes} evaluated using 
 the ESS method for structure functions 
 calculated in real space. The vertical lines indicate the scaling region 
 used to evaluate the scaling exponent, $\Sigma_n$.}
 \label{fig:ESS_local_slopes}
\end{figure}

 Figure \ref{fig:local_slopes} shows the calculated standard local slopes for four
 different Reynolds numbers $\Rl=42$, 64, 101 and 177.
 A plateau would indicate a constant exponent, that is a scaling region, but 
the figure does not show the formation of plateaux for these Reynolds numbers,
 implying that there is no scaling region.
If, in contrast the generalized structure functions are used to evaluate the 
local slopes $\Sigma_n$ as defined in \eqref{eq:sigma}, scaling regions are 
obtained and thus the ESS scaling exponents
can be measured. As shown in Fig.~\ref{fig:ESS_local_slopes}, extended plateaux 
for each order of generalized structure function are observed, which 
reach well into the dissipation scales where we do not expect power-law 
behavior.
Since $S_n \sim r^n$ as $r \to 0$, we note that $\Sigma_n(r) \to 
  n/3$ as $r \to 0$. It should be borne in mind that this K41-type behavior 
is an artefact of ESS, as has been pointed out before \cite{Benzi95,Barenblatt99,Benzi99}.
ESS exponents obtained from our calculations are shown to be consistent 
with relevant results from the literature in Table \ref{tbl:ESS_exponents}. 

\begin{table}[tb!]
 \begin{center}
  \begin{ruledtabular}
   \begin{tabular}{cccccccc}
   $R_\lambda$ & $\Sigma_2$ & $\Sigma_4$ & $\Sigma_5$ & $\Sigma_6$ & 
   $\Sigma_7$
   & $\Sigma_8$ & Source \\
   \hline\hline
  & 0.667 & 1.333 & 1.667 & 2.000 & 2.333 & 2.667 & K41 theory \\
   \hline
   42.5 & 0.690 & 1.287 & 1.551 & 1.796 & 2.023 & 2.234 & 
   \multirow{4}{*}{Our DNS, $\Sigma_n$} \\
   64.2 & 0.692 & 1.284 & 1.544 & 1.783 & 2.002 & 2.201 & \\
   101.3 & 0.692 & 1.283 & 1.544 & 1.785 & 2.008 & 2.215 & \\
   176.9 & 0.694 & 1.279 & 1.533 & 1.762 & 1.967 & 2.151 & \\
   \hline
   70 & 0.690 & 1.288 & 1.555 & 1.804 & 2.037 & 2.254 &
    \multirow{2}{*}{DNS \cite{Fukayama00}, $\Sigma_n$} \\
   125 & 0.692 & 1.284 & 1.546 & 1.788 & 2.011 & 2.217 &   \\
   \hline
   381 & 0.709 & 1.30 & 1.56 & 1.79 & 1.99 & 2.18 &
    \multirow{2}{*}{DNS \cite{Gotoh02}, $\zeta_n$} \\
   460 & 0.701 & 1.29 & 1.54 & 1.77 & 1.98 & 2.17 & \\
   \hline
   800 & 0.70 & 1.28 & 1.54 & 1.78 & 2.00 & 2.23 & Exp. \cite{Benzi95},
    $\Sigma_n$ \\
  \end{tabular}
  \end{ruledtabular}
 \end{center}
 \caption[Measurement of the scaling exponents using extended 
 self-similarity.]{Measurement of the scaling exponents from our DNS data 
 using ESS. For purposes of comparison, we also show the values predicted 
 from dimensional analysis (referred to as `K41 theory'), along with results 
 from Fukayama \etal\ \cite{Fukayama00}, Gotoh \etal\ \cite{Gotoh02} and 
 Benzi \etal\ \cite{Benzi95}. 
}
 \label{tbl:ESS_exponents}
\end{table}

\section{Spectral methods}
\label{sec:spectral}
The use of spectral methods to calculate the structure functions is known in 
the literature through the work by Bos \etal\ \cite{Bos12}, Tchoufag \etal\ \cite{Tchoufag12} 
and Qian \cite{Qian97,Qian99}. We will now briefly explain the reason
for doing it this way. 

Since the calculation of correlation and structure functions in real space 
requires a convolution in which the correlation of each site with every 
other (longitudinal) site needs to be measured, moving to large lattice sizes in 
order to reach larger Reynolds numbers results in a significant increase in 
computational workload. In addition, the number of realizations required to 
generate the ensemble takes both longer to produce and occupies substantially more 
storage space. In order to reach higher Reynolds numbers we have therefore 
used spectral expressions for the correlation and structure 
functions. Real-space quantities are then calculated by Fourier-transforming 
the appropriate spectral density. For example, the 
two-point correlation tensor may be found using
\begin{equation}
 C_{\alpha\beta}(\vec{r}) = \int d^3k\ \left\langle u_\alpha(\vec{k}) 
 u^*_\beta(\vec{k}) \right\rangle\ e^{i\vec{k}\cdot\vec{x}} \ .
\end{equation}
The assumption of isotropy then allows the transformation for the 
calculation of the isotropic correlation function to be reduced to
\begin{equation}
 C(r) = \tfrac{1}{2} C_{\alpha\alpha}(\vec{r}) = \int dk\ E(k) 
 \frac{\sin{kr}}{kr} \ .
\end{equation}
The real space correlation tensor $C_{\alpha \beta}(\vec{r})$, when written as a 
spatial average instead of an ensemble average (assuming ergodicity, as is the usual practice), 
is a convolution 
\beq
C_{\alpha \beta}(\vec{r})=\lim_{V \to \infty} \frac{1}{V} \int_V d\vec{x} \ 
u_{\alpha}(\vec{x})u_{\beta}(\vec{x}+\vec{r}) \ ,
\eeq 
of the velocity components $u_{\alpha}$ and $u_\beta$.
In a similar manner to the standard pseudospectral DNS technique, that is 
switching to real space in order to calculate the non-linear term and to 
avoid the convolution in Fourier space \cite{Yoffe12}, this approach replaces 
the convolution in real space with a Fourier transform of \emph{local} 
Fourier-space spectra. Furthermore, the shell-averaged spectra require 
substantially less in the way of storage and processing capabilities than 
real-space ensembles.

The derivations of the spectral representation of the second- and third-order structure functions 
are given in Appendix \ref{app:spec_rep}, leading to 
 \begin{equation}
  S_2(r) = 4\int_0^{\infty}dk\ E(k) a(kr) \ , 
 \end{equation}
and 
 \begin{equation}
  S_3(r) = 6C_{LL,L}(r)=12 \int_0^{\infty}dk\ \frac{T(k)}{k^2}\ \frac{\partial a(kr)}{\partial r} \ ,
 \end{equation}
where
\beq
a(x) = \frac{1}{3}   - \frac{\sin{x} - x\cos{x}}{x^3} \ ,
\eeq
in agreement with Bos \etal\ \cite{Bos12}.
The local slopes $\zeta_n$ can now be found by taking derivatives of the spectral
forms for the structure functions\footnote{Currently this is only possible for
$S_2(r)$ and $S_3(r)$, as the spectral expressions for the higher order structure
functions have not been derived yet.}, as shown in further detail in Appendix \ref{app:spec_rep}.

The spectral approach has the consequence that we are now evaluating the \emph{conventional} 
structure functions $S_n(r)$, 
rather than the \emph{generalized} structure functions, $G_n(r)$, as commonly used (including 
by us) for ESS. Before proceeding to the calculation of the scaling exponents, 
we will now discuss the effects of finite forcing and the calculation of viscous corrections 
to the second- and third-order structure functions, 
and subsequently validate the pseudospectral approach by comparing results for the spectrally obtained 
second- and third-order structure functions to corresponding real-space results.

\subsection{Effects of finite forcing on the structure functions}
The pseudospectral method can also be used to calculate the corrections due to finite 
forcing on the structure functions. The second- and third-order structure functions are related
by energy conservation, that is in real space by the K\'{a}rm\'{a}n-Howarth equation  
 \beq
  -\frac{3}{2}\frac{\partial U^2}{\partial 
  t} = -\frac{3}{4} \frac{\partial S_2}{\partial t} - \frac{1}{4r^4} 
  \frac{\partial}{\partial r} \Big( r^4 S_3 \Big) + \frac{3\nu_0}{2r^4} 
  \frac{\partial}{\partial r} \left( r^4 \frac{\partial S_2}{\partial r} 
  \right) \ ,
 \eeq
 where $(3/2) \partial U^2/\partial t = \partial E/\partial t = -\varepsilon_D$, 
with $\vep_D$ denoting the decay rate.
For decaying turbulence the decay rate equals the dissipation rate $\vep$, 
that is $\varepsilon_D = \varepsilon$.  For stationary turbulence $\vep_D=0$, 
and the dissipation rate then has to be equal to the energy input rate, that is 
$\varepsilon = \varepsilon_W$, where $\vep_W$ denotes the energy input rate, and 
the time-dependent $S_2$ term vanishes.
There is, however, a further complication. Replacing the dissipation rate $\vep$ 
with the energy input rate $\vep_W$, we see that the dissipation rate is
acting as an input. In the K41 theory, we have an equivalence between the
inertial transfer, dissipation, and input rates, in the infinite Reynolds
number limit. However, in using $\vep$ ($= \vep_W$) as
the input, it has been implicitly assumed that the forcing does not 
depend on the scale $r$. This is in general not the case, 
and as such some method for accounting for the effects of (finite) forcing
(\emph{i.e.} $Re < \infty$) must be introduced.
Various proposals have been put forward, mainly through the inclusion of corrections
to the KHE. These are discussed in Appendix \ref{app:strf_forcing}. An alternative treatment containing the 
exact energy input is given in the following section. 

\subsubsection{The KHE derived from the Fourier space energy balance}
\label{sec:new_KHE}

Instead of including a correction term in the KHE for forced turbulence, the effect of 
finite forcing can be calculated exactly using the spectral approach.
We begin with the energy balance equation in spectral space (nowadays referred to as the Lin
equation \cite{Sagaut08,McComb14a}) 
 \begin{equation}
 \label{eq:Lin}
  \frac{\partial E(k,t)}{\partial t} = T(k,t) - 2\nu_0 k^2 E(k,t) + W(k,t) \ ,
 \end{equation}
 where $W(k,t)$ is the work spectrum of the stirring forces and thus
contains the relevant information about the forcing. 
In order to obtain the energy balance equation in real space including the effects 
of (finite) forcing, we assume isotropy, take the Fourier transform of the Lin equation
and use the definitions of the structure functions $S_2$ and $S_3$ 
in order to obtain the energy balance equation in real space
(the KHE) relating $S_2$ and $S_3$ 
 \begin{equation}
  \label{eq:KHE_struct}
  \frac{\partial U^2}{\partial t} - \frac{1}{2}\frac{\partial 
  S_2(r)}{\partial t} = \frac{1}{6r^4} \frac{\partial}{\partial r} \Big( r^4 
  S_3(r) \Big) - \frac{\nu_0}{r^4} \frac{\partial}{\partial r} \left( r^4 
  \frac{\partial S_2(r)}{\partial r} \right) + \frac{2}{3} I(r,t) \ ,
 \end{equation}
 where the input term $I(r,t)$ is defined as
 \begin{equation}
  I(r,t) = \frac{3}{r^3}\int_0^r dy\ y^2 W(y,t) \ ,
 \end{equation}
and $W(y,t)$ is the (three-dimensional) Fourier transform of the work spectrum $W(k,t)$.
A detailed derivation of this equation from the Lin equation \eqref{eq:Lin} can be found in Appendix 
\ref{app:new_KHE}.

In order to write \eqref{eq:KHE_struct} in terms of energy loss, we multiply by $-3/2$ on both sides and obtain  
 \begin{equation}
  \varepsilon_D = - \frac{3}{4}\frac{\partial S_2(r)}{\partial t} 
  -\frac{1}{4r^4} \frac{\partial}{\partial r} \Big( r^4 S_3(r) \Big) + 
  \frac{3\nu_0}{2r^4} \frac{\partial}{\partial r} \left( r^4 \frac{\partial 
  S_2(r)}{\partial r} \right) - I(r,t) \ .
 \end{equation}
This is now the general form of the KHE including the effects of general, unspecified forcing. 
For the case of free decay, $\varepsilon_D = \varepsilon$ and the input term $I(r,t) = 0$, 
which leads to the well-known KHE for free decay. On the other hand, for the 
stationary case
$\varepsilon_D = 0$, $\partial S_2/\partial t = 0$ and the input term is independent of time $I(r,t) = I(r)$, thus 
 \begin{equation}
 \label{eq:KHE}
  I(r) =  
  -\frac{1}{4r^4} \frac{\partial}{\partial r} \Big( r^4 S_3(r) \Big) + 
  \frac{3\nu_0}{2r^4} \frac{\partial}{\partial r} \left( r^4 \frac{\partial 
  S_2(r)}{\partial r} \right) \ .
 \end{equation}
This derivation shows that the dissipation rate should \emph{not} be present in the 
KHE for stationary turbulence with finite forcing. Only for the limit of 
$\delta(\vec{k})$-forcing do we obtain $I(r)=\vep_W = \vep$ and hence recover 
the dissipation rate as an \emph{input} term. 

In contrast to previous attempts to include the effects of forcing into the 
KHE, we do not approximate the work term.
Instead, we use full information of this term as supplied by the work
spectrum. In this way
an explicit form for the actual energy input by the stirring forces can be calculated,
as we shall now show.
By integrating \eqref{eq:KHE} with respect to $r$ one obtains
  \begin{equation}
   S_3(r) = X(r) + 6\nu_0 \frac{\partial S_2}{\partial r} \ ,
  \end{equation}
with the input due to finite forcing
\beq 
  X(r) = - \frac{4}{r^4} \int_0^r dy\ y^4 I(y)
\eeq
evaluated using the spectral method 
  \begin{equation}
  \label{eq:finite_forcing}
   X(r) = -12r\int_0^{\infty}dk\ W(k)\ \left[ \frac{3\sin{kr} - 3kr\cos{kr} - 
   (kr)^2\sin{kr}}{(kr)^5} \right] \ .
  \end{equation}
Note that $X(r)$ is not a correction to K41, as used in previous studies.
Instead, it replaces the erroneous use of the dissipation rate and contains all the information of the energy input at all scales. In the limit of $\delta(\vec{k})$-forcing, $I(y) = \vep_W=\vep$, such that $X(r) 
  = -4\varepsilon r/5$, giving K41 in the infinite Reynolds number limit. 

\subsection{Real-space \emph{versus} spectral structure functions}
\begin{figure}
 \begin{center}
  \includegraphics[width=0.8\textwidth]{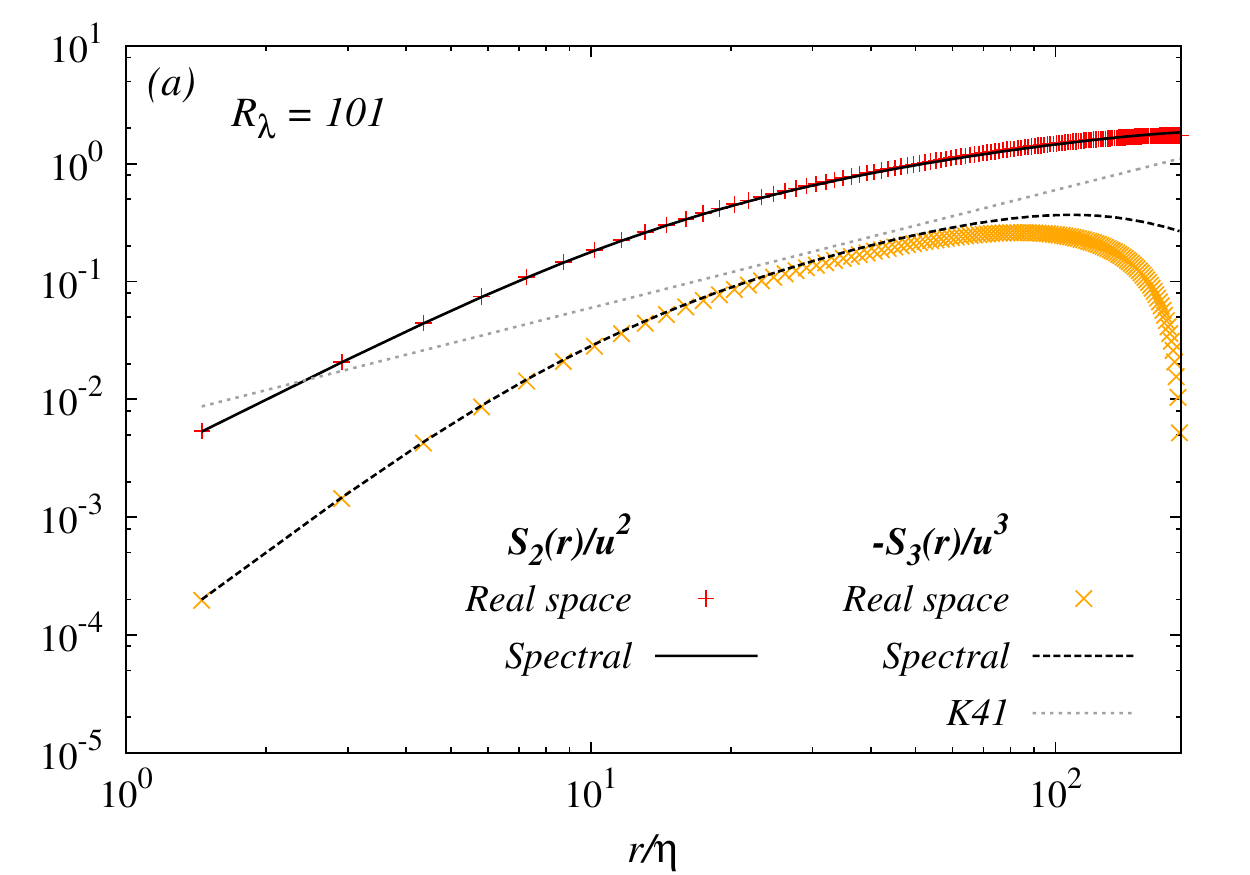}
  \includegraphics[width=0.8\textwidth]{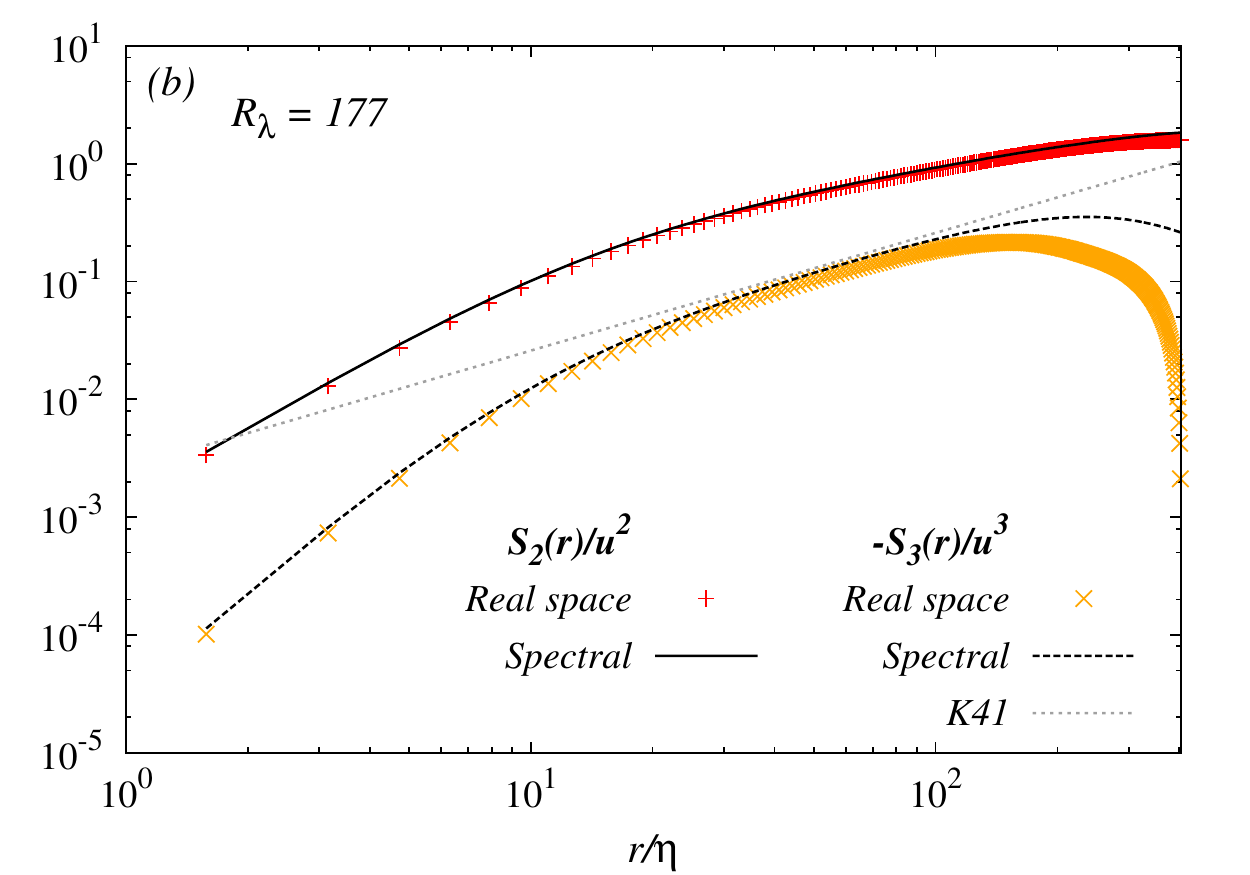}
 \end{center}
 \caption{(Color online) Comparison of second- and third-order (scaled) structure functions 
 calculated from the real-space ensembles and from energy and transfer spectra 
for two different Reynolds numbers. The agreement for $S_2$ is good at all 
scales, while for $S_3$ the results from real-space and spectral space 
calculations diverge at the 
large scales.}
 \label{fig:sf_comparison}
\end{figure}

\begin{figure}
 \begin{center}
  \includegraphics[width=0.8\textwidth]{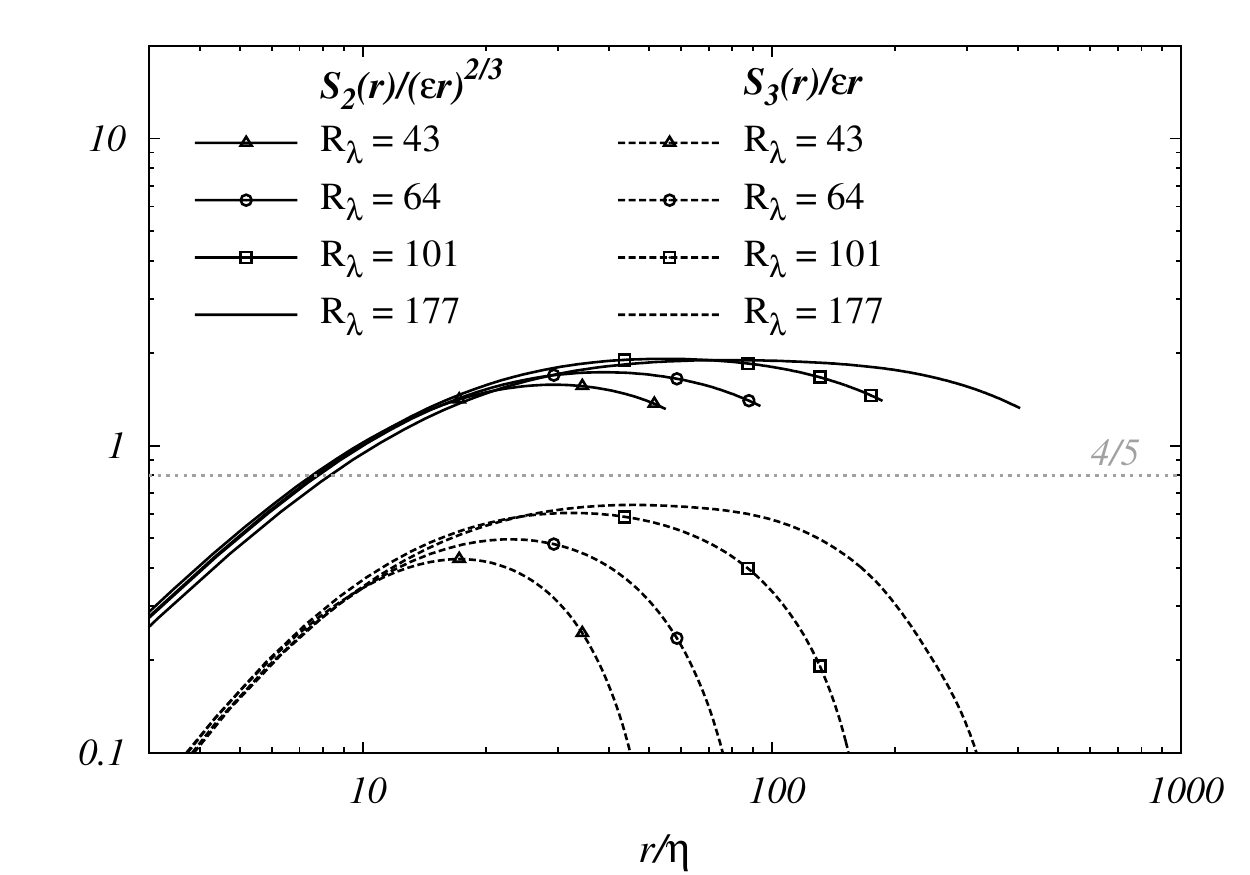}
 \end{center}
 \caption{Second- and third-order (scaled) structure functions, calculated 
 from the real-space ensembles. This figure should be compared to 
 Fig.~\ref{fig:sf_spectral}. The comparison shows how the real-space results
 can be extended to higher Reynolds numbers using the spectral method.}
 \label{fig:sf_realspace}
\end{figure}

Results from pseudospectral calculations of the structure functions were compared 
to real space ensemble averaged results for $R_\lambda = 101$ and $\Rl=177$. 
As can be seen in Fig.~\ref{fig:sf_comparison}, agreement for $S_2(r)$ is very good for all $r$.
For $S_3(r)$ we observe good agreement for small $r$, but the curves diverge at large $r$.
This could be due to DNS data being periodic in $L_\text{box} = 2\pi$. 
Since $S_3(r)$ is an odd function of $r$, it must go to zero in the center of the domain.
The pseudospectral method, however, involves a (weighted) superposition of damped 
oscillating functions which does not necessarily require that $S_3(L_\text{box}/2) = 0$.
Figure \ref{fig:sf_realspace} shows the compensated second- and third-order 
structure function calculated from real-space ensembles for the Taylor-scale
Reynolds 
number range $43 \leqslant \Rl \leqslant 177$. This may be compared to
figure 3(b) in the paper by Ishihara, Gotoh and Kaneda \cite{Ishihara09}.

\subsection{Spectral calculation from DNS}
\begin{figure}
 \begin{center}
  \includegraphics[width=0.8\textwidth]{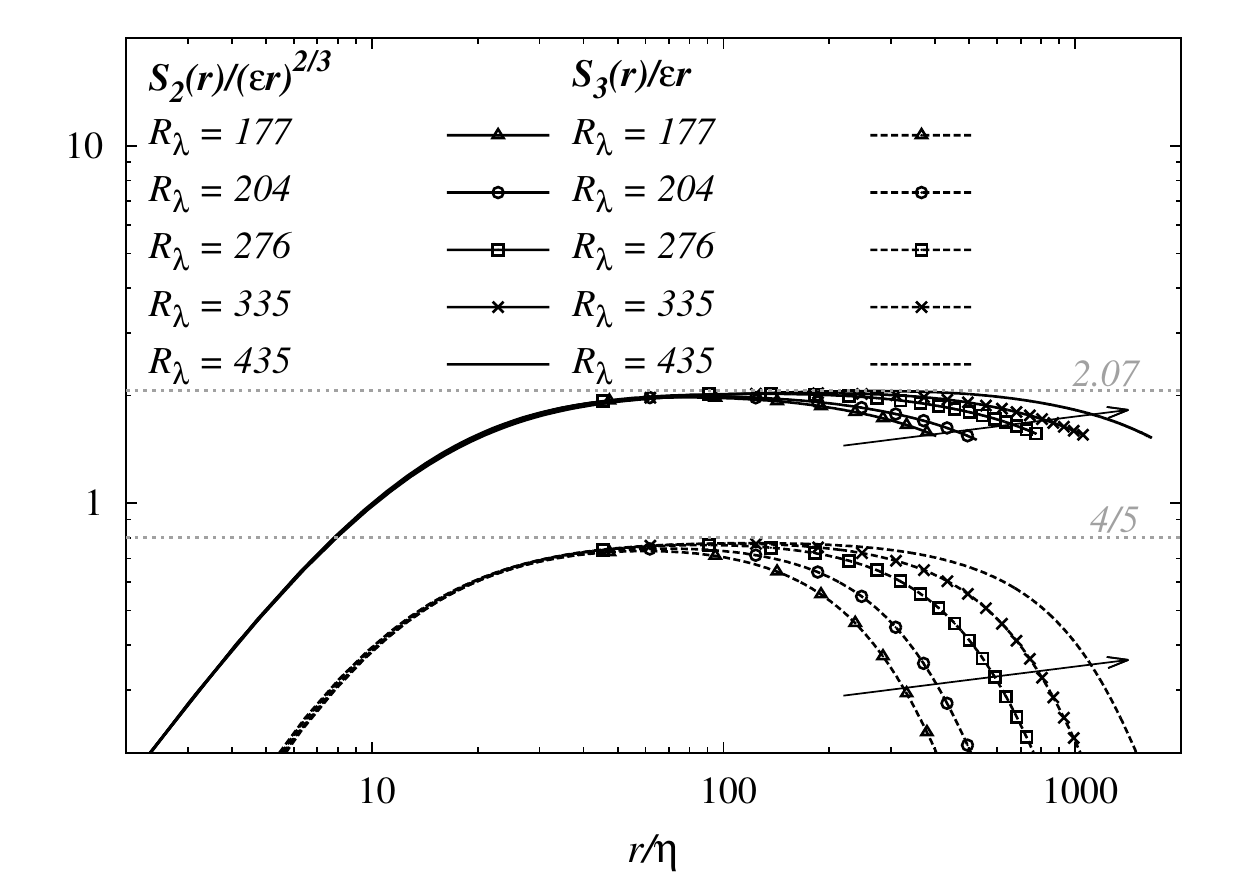}
 \end{center}
 \caption{Compensated second- and third-order (scaled) structure functions, calculated 
 from energy and transfer spectra. The horizontal lines indicate K41 scaling, where the 
upper horizontal line indicates our measured value $C_{K,S_2}=2.07$ 
for the prefactor $C_{K,S_2}$ of the second-order structure function.
The arrows indicate the direction of increasing Reynolds number, 
and one observes the formation of longer plateaux for both $S_2$ and $S_3$ with increasing 
Reynolds number indicating an approach to K41 scaling. }
 \label{fig:sf_spectral}
\end{figure}

The pseudospectral method was used to calculate structure functions for $\Rl=101$, $113$ 
and also for the higher Reynolds number range $177 \leqslant \Rl \leqslant 435$,
which can be seen in Fig.~\ref{fig:sf_spectral}, where the arrows indicate the direction 
of increasing Reynolds number.
The lower horizontal dotted line in the picture indicates K41 scaling for the third-order 
structure function, while the upper horizontal line indicates our measured value $C_{K,S_2}=2.07$ 
for the prefactor $C_{K,S_2}$ of the second-order structure function. The prefactor 
$C_{K,S_2}$ is related to the prefactor $C_{K,E_{LL}}$ of the longitudinal energy spectrum
$E_{LL}(k)$ by $C_{K,S_2} = 4.02 C_{K,E_{LL}}$ \cite{Monin75}, which has been measured by
Sreenivasan \cite{Sreenivasan95} to be $C_{K,E_{LL}}=0.53 \pm 0.055$. This results in 
$C_{K,S_2}=2.13 \pm 0.22$, compared to our measured value of $C_{K,S_2}=2.07$. 
Comparison of Figs.~\ref{fig:sf_realspace} and \ref{fig:sf_spectral} shows how the spectrally 
calculated results for the second- and third-order structure functions extend the real-space 
results to higher Reynolds numbers.   

\begin{figure}
 \begin{center}
  \includegraphics[width=0.8\textwidth]{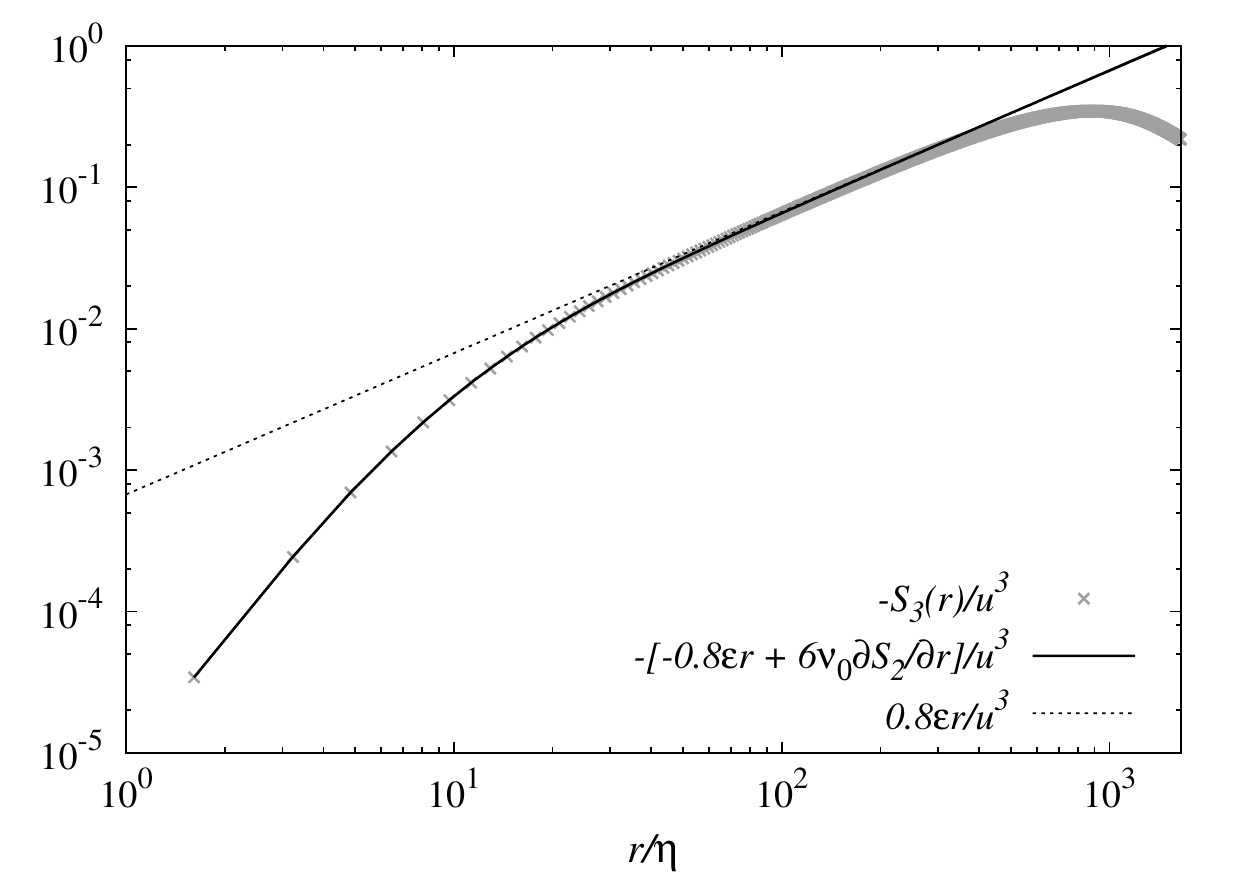}
 \end{center}
 \caption{Evaluation of the viscous correction to the `four-fifths'-law for
 the third-order structure function using the spectral method, with $\Rl=435.2$. The correction can be seen to account 
 for the difference between the `four-fifths'-law and DNS data at small scales.}
 \label{fig:sf_visc_correction}
\end{figure}

\begin{figure}
 \begin{center}
   \includegraphics[width=0.8\textwidth]{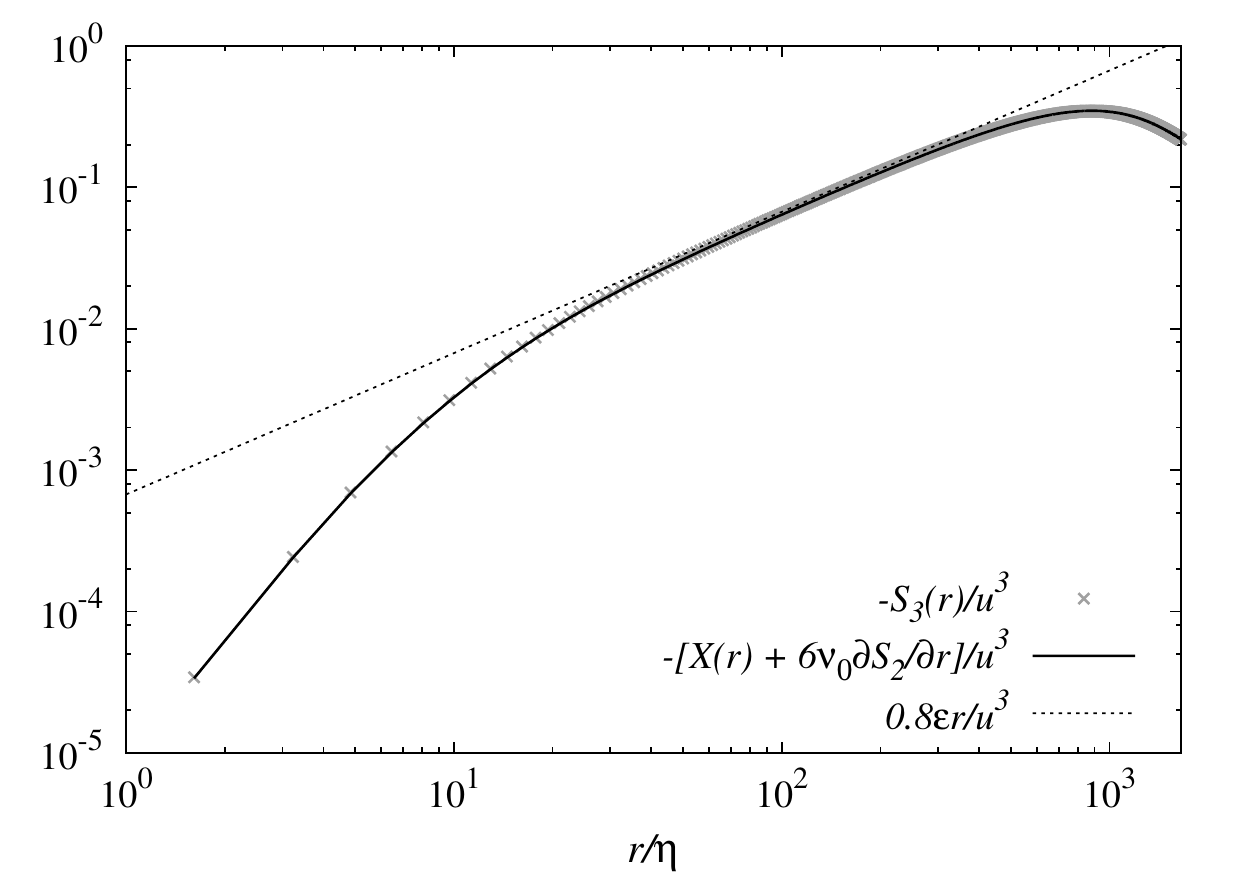}
 \end{center}
 \caption{Evaluation of the viscous correction to the input $X(r)$ (in place
 of the `four-fifths'-law) for the third-order structure
 function. Calculations were performed using the spectral method, with
 $\Rl=435.2$. The contributions can be seen to account for the differences
 between the `four-fifths'-law and DNS data at all scales.}
 \label{fig:sf_force_correction}
\end{figure}

The viscous correction to the `four-fifths'-law and to the exact input
contribution $X(r)$ calculated by the spectral method are shown in Fig.~\ref{fig:sf_visc_correction} 
and Fig.~\ref{fig:sf_force_correction}, respectively, with the third-order
structure function plotted for comparison. This is presented for our highest resolved simulation at $\Rl=435.2$.
Together with the viscous correction, the input $X(r)$ given in
\eqref{eq:finite_forcing} can be seen to account for differences between the
third-order structure function and the `four-fifths'-law at all scales, as
can be seen in Fig.~\ref{fig:sf_force_correction}. 
In contrast, Fig.~\ref{fig:sf_visc_correction} shows that the viscous correction alone 
only accounts for the difference between DNS data for the third-order
structure function and K41 at the small scales, as expected, 
since at scales much smaller than the forcing scale the system becomes insensitive to 
the details of the (large-scale) forcing.
   
\section{A new scaling exponent}
\label{sec:new_exponent}
We now arrive at our proposal to
introduce a new local-scaling exponent $\Gamma_n$, which can be used
to determine the $\zeta_n$. We work with $S_n(r)$ and
consider the quantity $\lvert S_n(r)/S_3(r)  \rvert$.   In this
procedure, the exponent $\Gamma_n$ is defined by
\begin{equation}
 \left\lvert \frac{S_n(r)}{S_3(r)} \right\rvert \sim r^{\Gamma_n} \ ,
 \qquad\text{where}\qquad \Gamma_n = \zeta_n - \zeta_3 \ .
\end{equation}
The definition of $\Gamma_n$ is motivated by a long-established
technique in experimental physics, where the effective experimental
error can be reduced by plotting the ratio of two dependent variables:
see \emph{e.g.}~Chapter 3 in the well-known book by Bevington and 
Robinson on data analysis \cite{Bevington03}. Of course this does not work in
all cases, but only where the quantities are positively correlated, and
we have verified that this is the case for $S_2$ and $\lvert S_3 \rvert$.

The error in the measurement of the $n^{th}$-order structure function can be
 expressed as
   \begin{equation}
    S_n(r) = \Big[ 1 + \epsilon_n(r) + O(\epsilon_n^2) \Big] 
    \overline{S}_n(r) \ ,
   \end{equation}
   where $\overline{S}_n(r)$ is the `true' value and $\epsilon_n(r)$ a 
   measurement of the systematic error and considered small.
   Hence if $\epsilon_n(r) \sim \epsilon(r)$, then the ratio 
   $\left\lvert \frac{S_n(r)}{S_3(r)} \right\rvert$  has an 
   error proportional to the second-order of a small quantity,
   \begin{equation}
    \frac{S_n(r)}{S_3(r)} = \frac{\overline{S}_n(r)}{\overline{S}_3(r)} 
    \Big[ 1 - \epsilon^2(r) + O(\epsilon^3) \Big] \ .
   \end{equation}
  For illustration purposes we assumed here that $S_n$ and $S_3$ were 
  perfectly correlated, note that for imperfect correlation there is still
  a reduction in error. 
  The local slope now is found by considering $\Gamma_2(r) = 
  \zeta_2(r) - \zeta_3(r)$. 
  By once again assuming that $\zeta_3(r) = 1$, the local slope for the 
  second-order structure function is found as $\Gamma_2(r) + 1 = 
  \zeta_2(r)$.

\subsection{Scaling exponents from spectra}

\begin{figure}[tbp]
 \begin{center}
  \includegraphics[width=0.8\textwidth]{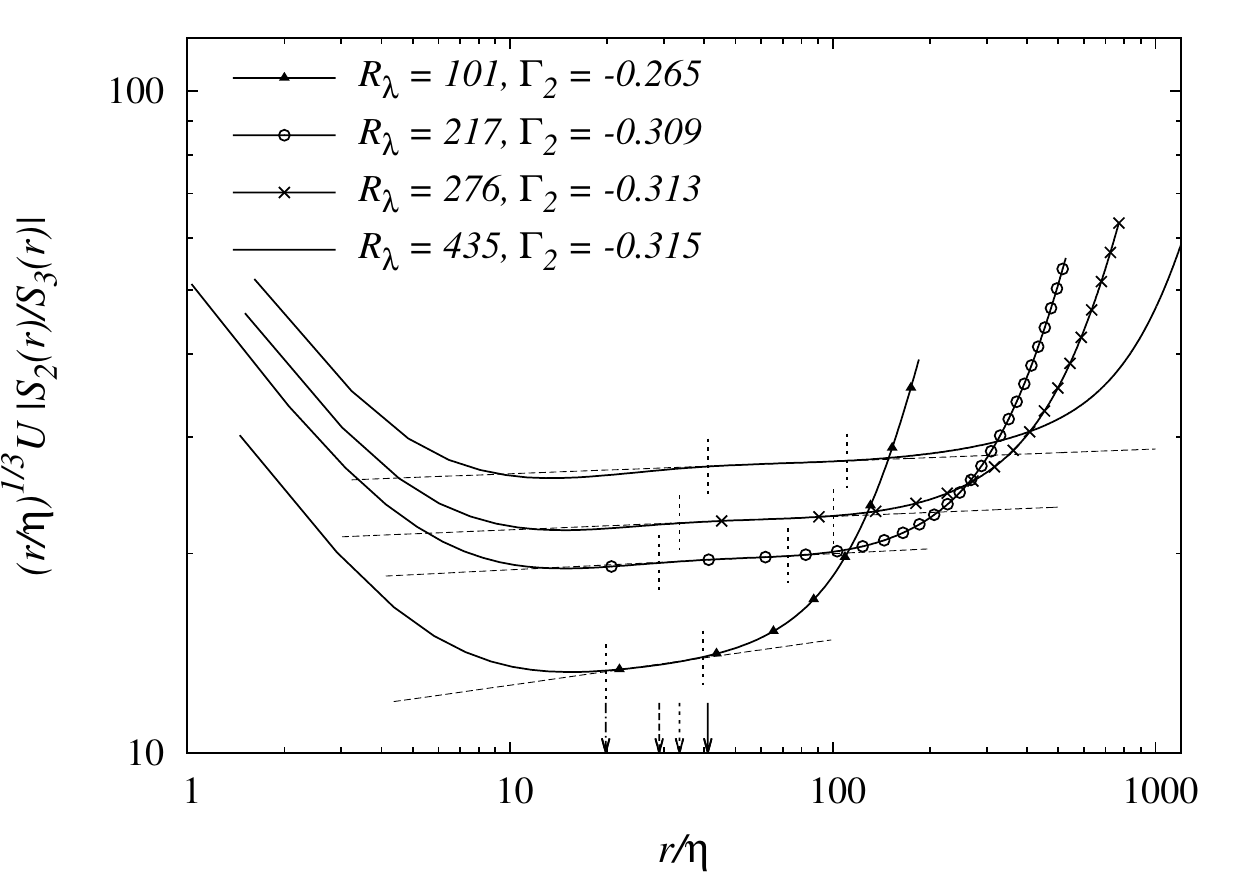}
 \end{center}
 \caption{The compensated ratio $(r/\eta)^{1/3}U\lvert 
 S_2(r)/S_3(r)\rvert$ plotted against $r$ and scaled on the
 dissipation scale, $\eta$.  K41 scaling would correspond to a
 plateau. Arrows indicate $\lambda/\eta$, while the vertical dotted lines
 show the region used to fit each exponent. Note that the measured compensated
 slopes become flatter with increasing Reynolds number, pointing towards the deviation 
from Kolmogorov scaling being a finite Reynolds number effect.}
 \label{fig:wdmss}
\end{figure}

The new scaling exponent $\Gamma_n$, based on the conventional (as
opposed to the \emph{generalized}) structure functions
$S_n(r)$, is compatible 
with spectral methods and has been tested for the case $n=2$ in Fig.~\ref{fig:wdmss}.
The dimensionless quantity $U\lvert S_2(r)/S_3(r) \rvert$, where $U$ is
the rms velocity,  is plotted against $r/\eta$, for three values of
$R_\lambda$. Note that, since K41 predicts $\Gamma_2 = -1/3$, we have
plotted a compensated form, in which we multiply the ratio by
$(r/\eta)^{1/3}$, such that K41 scaling would correspond to a plateau. From
the figure, we can see a trend towards K41 scaling as the Reynolds
number is increased.
Note that this figure also illustrates the ranges used to find
values for our new exponent $\Gamma_2$, for the following cases.
$\Gamma_2$ was fitted to the ranges $\lambda < r < c\lambda$, with $c =
2.0, 2.5, 2.6$, and 2.7 for $R_\lambda = 101, 217, 276$ and 435, respectively.

\begin{figure} [tbp]
 \begin{center}
  \includegraphics[width=0.8\textwidth]{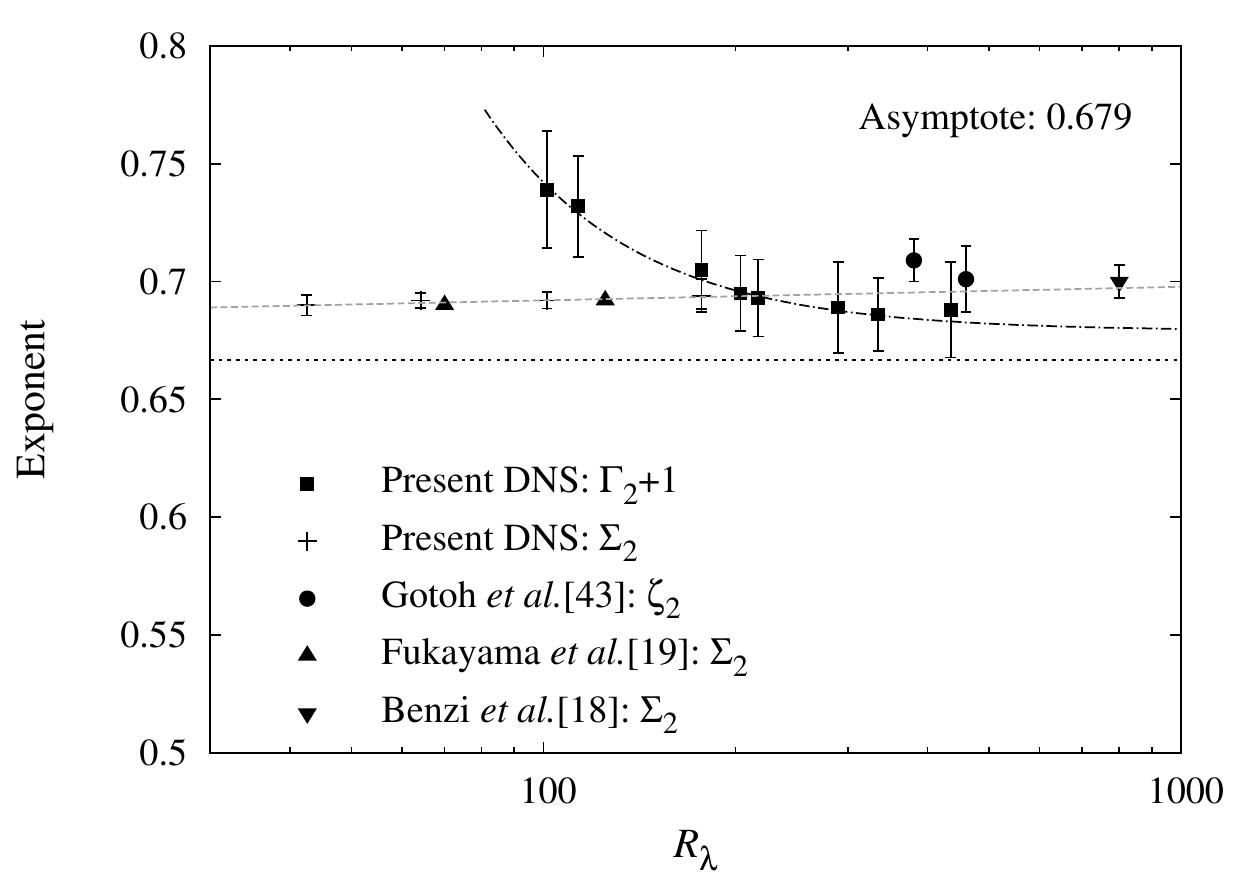}
 \end{center}
\caption{Dependence of our new exponent $\zeta_2 = \Gamma_2 
+ 1$ on Reynolds number, compared to measured ESS exponents denoted
by $\Sigma_2$ from
Fukayama \etal\ \cite{Fukayama00}, Benzi \etal\ \cite{Benzi95} and our DNS, 
as well as results for $\zeta_2$ by Gotoh \etal\ \cite{Gotoh02}. 
The horizontal line indicates the K41 value
of 2/3. The dash-dot line indicates the fit to $\Gamma_2 + 1$, while the
dashed line shows a fit to $\Sigma_2$ using our results and those of
Fukayama \etal\ \cite{Fukayama00}.}
 \label{fig:exponents_summary}
\end{figure}

Figure \ref{fig:exponents_summary} summarizes the comparison between our
results for our new method of determining the second-order exponent and
those based on ESS (our own and others \cite{Benzi95,Fukayama00}) or on
direct measurement \cite{Gotoh02}, in terms of their overall dependence
on the Taylor-Reynolds number. In order to establish the form of the
dependence of the exponents on Reynolds number, we fitted curves to the
data points using the nonlinear-least-squares Marquardt-Levenberg
algorithm, with the error quoted being one standard error. Using the data 
obtained from our new method, we fitted a curve $\Gamma_2 + 1 = A + B R_\lambda^p$, 
to find the asymptotic value $A = 0.679 \pm 0.013$, which is consistent
with the deviations from K41 scaling being finite Reynolds number effects. 
In order to compare this result with results obtained by the ESS method, 
we fitted the curve $\Sigma_2= C + D R_\lambda^q$ to our own data plus that of 
Fukayama \etal\ \cite{Fukayama00}. Evidently the two fitted curves 
show very different trends, with results for $\Sigma_2$ increasing with increasing
Reynolds number, whereas $\zeta_2 = \Gamma_2 + 1$ decreases and
approaches 2/3 (within one standard error) as $R_\lambda$ increases.

It should be emphasized that with both methods, that is ESS and our new method,
 it is necessary to take
$\zeta_3=1$ in the inertial range, in order to obtain the inertial-range
value of either $\Sigma_2 = \zeta_2$ (by ESS) or $\Gamma_2 = \zeta_2 -1$
(our new method). For this reason, we plot $\Gamma_2+1$, rather than
$\Gamma_2$ in Fig.~\ref{fig:exponents_summary}.  
An obvious difference between our proposed method and ESS is apparent  
as $r \to 0$. This is readily understood in terms of the regularity 
condition for the velocity field, which leads to $S_n(r) \sim r^n$ as $r \to 0$
\cite{Stolovitzky93,Sirovich94}. This yields $\Gamma_n(r) + 1 \to n -
2$, whereas ESS gives  $\Sigma_n(r) \to n/3$.

In this context it may be of interest to briefly discuss the experimental results 
of Mydlarski and Warhaft \cite{Mydlarski96}, who measured the exponent of the 
longitudinal energy spectrum for a range of Taylor-Reynolds numbers from 
$\Rl=50$ to $\Rl=473$, which is similar to the range of Taylor-Reynolds numbers studied
in the present paper. The authors found the inertial-range exponent $\alpha$ 
of the longitudinal energy spectrum $E_{LL}(k)$ 
to depend on $\Rl$ 
in the following way 
\beq
\alpha = -\frac{5}{3}\left(1+3.15 \Rl^{-2/3}\right) \ ,   
\eeq
hence $\alpha \to -5/3$ in the limit of infinite Reynolds number. 
Thus the results of Mydlarski and Warhaft support our result for the exponent of $S_2$, since 
$\alpha \to -5/3$ implies $\zeta_2 \to 2/3$ as $\Rl \to \infty$.
\section{Conclusions}

As we have said in the introduction, the point at issue is essentially `intermittency
corrections versus finite Reynolds number effects'. The former has
received much more attention; but, in recent years, there has been a
growing interest in studying finite Reynolds number effects,
experimentally and by DNS, for the case of $S_3$: see
\cite{Tchoufag12,Gotoh02,Antonia06} and references therein. (Although we
note that in this case the emphasis is on the prefactor rather
than the exponent.)

Our new result that  $\zeta_2 = \Gamma_2 +  1 \to 2/3$
is an indication that anomalous
values of $\zeta_2$ are due to finite Reynolds number effects, consistent
with the experimental results of Mydlarski \etal\, which point in the 
same direction.
Previously it had been suggested by Barenblatt \etal\ that ESS could be
interpreted in this way \cite{Barenblatt99}, but this was disputed by
Benzi \etal\ \cite{Benzi99}.

There is much remaining to be understood about these matters and we
suggest that our new method of analyzing data can help. It should, of
course, be noted that our use of $S_3$ (as evaluated by pseudospectral
methods) rather than $G_3$ (as used with ESS), may also be a factor in
our result. As a matter of
interest, we conclude by noting that our analysis could provide a
stimulus for further study of ESS and may lead to an understanding 
of the relationship between the two methods. It is also the case that 
the pseudospectral method could be used for the general study of higher-order structure
functions, but this awaits the derivation of the requisite Fourier
transforms.

\acknowledgments
The authors would like to thank Matthew Salewski, who read a first draft
and made many helpful suggestions. 
This work has made use of the resources provided by HECToR
\cite{hector}, made available through the Edinburgh
Compute and Data Facility (ECDF)
\cite{ecdf}.
A.~B. is supported by STFC, S.~R.~Y. and M.~F.~L. are funded by EPSRC.

\appendix 

\section{Derivation of the spectral representation of the structure functions}
\label{app:spec_rep}
The second order structure function can be derived from the  
relationship between the isotropic correlation function $C(r)$ and the energy 
  spectrum $E(k)$ where
 \begin{equation}
  \label{eq:C_E}
  C(r) = \int_0^{\infty} dk\ E(k)\ \frac{\sin{kr}}{kr} \ , 
 \end{equation}
and
 \begin{equation}
  E(k) = \frac{2}{\pi} \int_0^{\infty} dr\ C(r)\ kr\sin{kr} \ ,
 \end{equation}
as \emph{e.g.}~in Batchelor \cite{Batchelor53} equation (3.4.15) p49.
Equation \eqref{eq:C_E} can be used to derive a spectral expression for the 
longitudinal correlation function
$C_{LL}(r)$, since $C(r)$ and $C_{LL}(r)$ are related through
 \begin{equation}
  \label{eq:C_CLL}
  C(r) = \frac{1}{2r^2} \frac{\partial}{\partial r} \Big[ r^3 C_{LL}(r) \Big] \ ,  
 \end{equation}
which can be integrated to give
 \begin{equation}
C_{LL}(r) = \frac{2}{r^3} \int_0^r dy\   y^2 C(y) \ .
 \end{equation}
Now the integral over $y$ is done analytically to obtain
 \begin{equation}
  C_{LL}(r) = 2 \int_0^{\infty} dk\ E(k)\ \left[ \frac{\sin{kr} - kr\cos{kr}}{(kr)^3} 
  \right] \ ,
 \end{equation}
 as in Monin and Yaglom \cite{Monin75} vol.~2, equation (12.75). 
The spectral expression for the second-order structure function is now readily seen to be
 \begin{equation}
  S_2(r) = 2U^2 - 2C_{LL}(r) = 2\cdot\underbrace{\frac{2}{3}\int_0^{\infty} dk\ 
  E(k)}_{U^2} -\ 4 \int_0^{\infty} dk\ E(k)\ \left[ \frac{\sin{kr} - 
  kr\cos{kr}}{(kr)^3} \right] \ ,
 \end{equation}
 which can be written in a more concise form 
 \begin{equation}
  S_2(r) = 4\int_0^{\infty}dk\ E(k) a(kr) \ , 
 \end{equation}
using 
\beq
a(x) = \frac{1}{3}   - \frac{\sin{x} - x\cos{x}}{x^3} \ .
\eeq
Note that $a(0)=0$ since
 \begin{equation}
  \lim_{x \to 0} \frac{\sin{x} - x\cos{x}}{x^3} = \frac{1}{3} \ .
 \end{equation}

 Similarly, the spectral expression for the third-order structure function can be derived from 
 the relationship between the isotropic third-order correlation function and 
 the transfer spectrum
 \begin{equation}
  \label{eq:CLLL_T}
  \frac{1}{2} \left( 3 + r\frac{\partial}{\partial r} \right) \left( 
  \frac{\partial}{\partial r} + \frac{4}{r} \right) C_{LL,L}(r) = 
  \frac{1}{2r^2} \frac{\partial}{\partial r} \left[ \frac{1}{r} 
  \frac{\partial}{\partial r} \Big( r^4 C_{LL,L}(r) \Big) \right] = \int_0^{\infty}dk\ 
  T(k)\ \frac{\sin kr}{kr} \ ,
 \end{equation}
as in \emph{e.g.}~Batchelor \cite{Batchelor53} equation (5.5.14) p.~101, where 
Batchelor's $K(r)$ corresponds to \\
$(1/r^4) (\partial/\partial r) r^4 C_{LL,L}(r)$  
in our notation. After integrating by parts (with respect to $r$) one obtains 
 \begin{align}
  C_{LL,L}(r) &= 2r \int_0^{\infty}dk\ T(k)\ \left[ \frac{3\sin{kr} - 3kr\cos{kr} - 
  (kr)^2\sin{kr}}{(kr)^5} \right] \\
  &= 2 \int_0^{\infty}dk\ \frac{T(k)}{k^2}\ \frac{\partial a(kr)}{\partial r} \ ,
 \end{align}
 see \emph{e.g.}~Monin and Yaglom \cite{Monin75} vol.~2, equation (12.141$'''$). 
 The spectral expression for the third-order structure function follows directly
 \begin{equation}
  S_3(r) = 6C_{LL,L}(r)=12 \int_0^{\infty}dk\ \frac{T(k)}{k^2}\ \frac{\partial a(kr)}{\partial r} \ .
 \end{equation}

The local slopes $\zeta_1$ and $\zeta_2$ can now be found by taking derivatives of the spectral 
forms for the structure functions 
  \begin{align}
   \zeta_2(r) &= \frac{r}{S_2(r)} \frac{\partial S_2(r)}{\partial r} = 
   \frac{4r}{S_2(r)} \int_0^{\infty}dk\ E(k)\ \frac{\partial a(kr)}{\partial r} 
   \nonumber \\
   &= \frac{4}{S_2(r)} \int_0^{\infty}dk\ E(k)\ \left[ \frac{3\sin{kr} - 3kr\cos{kr} - 
   (kr)^2\sin{kr}}{(kr)^3} \right] \ ;\\
   \zeta_3(r) &= \frac{r}{S_3(r)} \frac{\partial S_3(r)}{\partial r} = 
   \frac{12r}{S_3(r)} \int_0^{\infty}dk\ \frac{T(k)}{k^2}\ \frac{\partial^2 
   a(kr)}{\partial r^2} \nonumber \\
   &= \frac{12r}{S_3(r)} \int_0^{\infty}dk\ T(k)\ \left[ \frac{\big( 5(kr)^2 - 12 
   \big) \sin{kr} - \big( (kr)^2 - 12 \big) kr\cos{kr}}{(kr)^5} \right] \ .
  \end{align}

\section{Previous attempts to include the effects of forcing in 
studies of the structure functions}
\label{app:strf_forcing}
   Gotoh \etal\ \cite{Gotoh02} studied a `generalized' KHE equation defined through their 
   equation (27), which is rewritten here as
   \begin{equation}
    \varepsilon = - \frac{1}{4r^4}\frac{\partial}{\partial r} \Big( r^4 S_3 
    \Big) + \frac{3\nu_0}{2r^4} \frac{\partial}{\partial r} \left( r^4 
    \frac{\partial S_2}{\partial r} \right) + \frac{3}{4r}I_G(r) \ ,
   \end{equation}
   with the input defined from their equation (28) in terms of our input term, 
   $I(r)$, as
   \begin{equation}
    I_G(r) = 4r\int_0^{\infty}dk\ W(k)\ \left[ \frac{1}{3} - 
    \frac{\sin{kr}-kr\cos{kr}}{(kr)^3} \right] = \frac{4r}{3} \Big( 
    \varepsilon_W - I(r) \Big) \ .
   \end{equation}
   Gotoh \etal\ 
   retain the dissipation rate in the KHE, despite its origin as $\partial 
   U^2/\partial t = 0$ for forced turbulence. They then find a correction in order
   to compensate for the retained dissipation rate, which also includes
   the work done by the stirring forces. Note that $\varepsilon = \varepsilon_W$ 
   cancels on both sides of the equation.
   The correction is then approximated, since the forcing is 
   confined to low wavenumbers
   \begin{equation}
    I_G(r) \simeq \frac{2}{15}\varepsilon_W K^2 r^3 \ , 
   \end{equation}
where 
   \begin{equation}
    K^2 = \frac{\int_0^{\infty}dk\ k^2 W(k)}{\int_0^{\infty}dk\ W(k)} = 
    \frac{1}{\varepsilon_W} \int_0^{\infty}dk\ k^2 W(k) \ .
   \end{equation}
   This approximation is plotted in figure 13 of \cite{Gotoh02}, and can be 
   seen to give good agreement to DNS data for most scales.
   It is also used to express $S_3(r)$ in terms of $S_2$ 
   \begin{equation}
    \label{eq:Got_corr}
    S_3(r) = -\frac{4\varepsilon r}{5} + 6\nu_0 \frac{\partial S_2}{\partial r} + \frac{2}{35}\varepsilon_W K^2 r^3 \ ,
   \end{equation}
   as shown in (their) figure 11. This approach has also been discussed in Kaneda \etal\ \cite{Kaneda08}.

  Sirovich \etal\ \cite{Sirovich94} use 
\begin{equation}
  \label{eq:Sir_corr} 
    S_3(r) = -\frac{4\varepsilon r}{5}
    + 6\nu_0 \frac{\partial S_2}{\partial r}
    + \frac{6}{r^4} \int_0^r dy\ y^4
    \big\langle \delta u_L(y) \delta f_L(y) \big\rangle \ ,
  \end{equation}
   where the longitudinal force increment is defined (in a similar manner to 
   $\delta u_L(r)$) as
   \begin{equation}
    \delta f_L(r) = \Big[ \vec{f}(\vec{x}+\vec{r}) - \vec{f}(\vec{x}) \Big] 
    \cdot \hat{\vec{r}} \ .
   \end{equation}
   This approach also retains the dissipation rate alongside a 
   correction term.
   The integral in \eqref{eq:Sir_corr} is approximated to give
   \begin{equation}
    S_3(r) \simeq -\frac{4\varepsilon r}{5} + 6\nu_0 \frac{\partial S_2}{\partial r} + \frac{2}{7}\varepsilon k_0^2 r^3 \ ,
   \end{equation}
   with $k_0$ the forcing wavenumber. This can be compared to \eqref{eq:Got_corr} for the result
   obtained by Gotoh \etal\ .

\section{Derivation of the KHE from the Fourier space energy balance}
\label{app:new_KHE}

In order to obtain the energy balance equation in real space including the effects 
of (finite) forcing, we assume isotropy and take the Fourier transform of the Lin equation \eqref{eq:Lin}
 \begin{align}
  \frac{\partial}{\partial t}\int_0^{\infty}& dk\ E(k,t)\ \frac{\sin kr}{kr} \nonumber \\ &= \int_0^{\infty}dk\ 
  T(k,t)\ \frac{\sin kr}{kr} - 2\nu_0 \int_0^{\infty}dk\ k^2 E(k,t)\ 
  \frac{\sin{kr}}{kr} + \int_0^{\infty}dk\ W(k,t)\ \frac{\sin kr}{kr} \ .
 \end{align}
For the work term we obtain
 \begin{equation}
  W(r,t) = \int_0^{\infty}dk\ W(k,t)\ \frac{\sin kr}{kr} = \frac{1}{2} \Big[ \langle 
  u_\alpha(\vec{x}) f_\alpha(\vec{x}+\vec{r}) \rangle + \langle 
  f_\alpha(\vec{x}) u_\alpha(\vec{x}+\vec{r}) \rangle \Big] \ .
 \end{equation}
For the dissipation term, note that
 \begin{equation}
  \left( \frac{2}{r} + \frac{\partial}{\partial r} \right) 
  \frac{\partial}{\partial r} \int_0^{\infty}dk\ E(k,t)\ \frac{\sin kr}{kr} 
= -\int_0^{\infty}
  dk\ k^2 E(k,t)\ \frac{\sin kr}{kr} \ ,
 \end{equation}
thus, using \eqref{eq:CLLL_T} and \eqref{eq:C_E}, we obtain for the energy balance equation in real space
 \begin{equation}
  \frac{\partial C(r)}{\partial t} = \frac{1}{2r^2} \frac{\partial}{\partial 
  r} \left[ \frac{1}{r} \frac{\partial}{\partial r} \Big( r^4 C_{LL,L}(r) 
  \Big) \right] + \frac{2\nu_0}{r^2} \frac{\partial}{\partial r} \left( r^2 
  \frac{\partial C(r)}{\partial r} \right) + W(r,t) \ .
 \end{equation}
 By using the relation \eqref{eq:C_CLL} between $C(r)$ and $C_{LL}(r)$, multiplying by $2r^2$ integrating 
 once over $r$ and finally dividing by $r^3$ we obtain the KHE in terms of the longitudinal
 correlation functions
 \begin{equation}
  \frac{\partial C_{LL}(r)}{\partial t} = \frac{1}{r^4} 
  \frac{\partial}{\partial r} \Big( r^4 C_{LL,L}(r) \Big) + 
  \frac{2\nu_0}{r^4} \frac{\partial}{\partial r} \left( r^4 \frac{\partial 
  C_{LL}(r)}{\partial r} \right) + \frac{2}{r^3}\int_0^r dy\ y^2 W(y,t) \ .
 \end{equation}

Now we can insert the definitions of the structure functions and arrive at 
the KHE relating $S_2$ and $S_3$ 
 \begin{equation}
  \frac{\partial U^2}{\partial t} - \frac{1}{2}\frac{\partial 
  S_2(r)}{\partial t} = \frac{1}{6r^4} \frac{\partial}{\partial r} \Big( r^4 
  S_3(r) \Big) - \frac{\nu_0}{r^4} \frac{\partial}{\partial r} \left( r^4 
  \frac{\partial S_2(r)}{\partial r} \right) + \frac{2}{3} I(r,t) \ ,
 \end{equation}
 where the input term $I(r,t)$ is defined as
 \begin{equation}
  I(r,t) = \frac{3}{r^3}\int_0^r dy\ y^2 W(y,t) \ .
 \end{equation}

\bibliographystyle{unsrt}
\bibliography{wdm_da}

\end{document}